%% file: main.tex
\documentclass[acmsmall,screen, nonacm]{acmart}
\makeatletter
\let\@authorsaddresses\@empty
\makeatother

\renewcommand\footnotetextcopyrightpermission[1]{} 
\AtBeginDocument{%
  \providecommand\BibTeX{{%
    \normalfont B\kern-0.5em{\scshape i\kern-0.25em b}\kern-0.8em\TeX}}}

\usepackage{siunitx}
\usepackage{booktabs}
\usepackage{multirow}
\usepackage{tabularx}
\usepackage{longtable}
\usepackage{hyperref}
\usepackage{balance}
\usepackage{xurl}
\usepackage[nameinlink]{cleveref}
\usepackage[official]{eurosym}

\newcolumntype{H}{>{\setbox0=\hbox\bgroup}c<{\egroup}@{}}
\newcolumntype{Z}{>{\setbox0=\hbox\bgroup}c<{\egroup}@{\hspace*{-\tabcolsep}}}

\usepackage[inline]{enumitem}

\usepackage{caption}
\usepackage{subcaption}
\hypersetup{%
  bookmarksnumbered,
  colorlinks = true,
  breaklinks=true,
  hypertexnames=false,
  pdfdisplaydoctitle=true, 
  pdfborder = {0 0 0},
  pdfstartview={FitH},
  pdflang={en-US},
  pdfkeywords={},
  pdftitle={52 Weeks Later: Attitudes Towards COVID-19 Apps for Different Purposes Over Time},
}

\newcommand\changed[1]{{{\color{black}#1}}}

\newcommand\changedcscw[1]{{{{#1}}}}
\newcommand\changedminor[1]{{{{#1}}}}

\definecolor{yellowfactor}{HTML}{D41159}
\newcommand{\factor}[1]{\textcolor{yellowfactor}{#1}}
\definecolor{bluefactor}{HTML}{1A85FF}
\newcommand{\factorrounds}[1]{\textcolor{bluefactor}{#1}}

\newcommand\old[1]{}

\newcommand{\ie}{i.\,e.}
\newcommand{\eg}{e.\,g.}

\begin{document}

\title[Attitudes Towards Different COVID-19 Apps Over Time]{52 Weeks Later: Attitudes Towards COVID-19 Apps for Different Purposes Over Time}

\author{Marvin Kowalewski}
\affiliation{%
  \institution{Ruhr University Bochum}
  \city{Bochum}
  \country{Germany}}
\email{marvin.kowalewski@rub.de}

\author{Christine Utz}
\affiliation{%
  \institution{CISPA Helmholtz Center for Information Security}
  \city{Saarbr\"{u}cken}
  \country{Germany}}
\email{christine.utz@cispa.de}

\author{Martin Degeling}
\affiliation{%
  \institution{Ruhr University Bochum}
  \city{Bochum}
  \country{Germany}}
\email{martin.degeling@rub.de}

\author{Theodor Schnitzler}
\affiliation{%
  \institution{Research Center Trustworthy Data Science and Security}
  \city{Dortmund}
  \country{Germany}}
\affiliation{%
  \institution{TU Dortmund University}
  \city{Dortmund}
  \country{Germany}}
\email{theodor.schnitzler@tu-dortmund.de}

\author{Franziska Herbert}
\affiliation{%
  \institution{Ruhr University Bochum}
  \city{Bochum}
  \country{Germany}}
\email{franziska.herbert@rub.de}

\author{Leonie Schaewitz}
\affiliation{%
  \institution{Ruhr University Bochum}
  \city{Bochum}
  \country{Germany}}
\email{leonie.schaewitz@rub.de}

\author{Florian M. Farke}
\affiliation{%
  \institution{Ruhr University Bochum}
  \city{Bochum}
  \country{Germany}}
\email{florian.farke@rub.de}

\author{Steffen Becker}
\affiliation{%
  \institution{Ruhr University Bochum}
  \city{Bochum}
  \country{Germany}}
\affiliation{%
  \institution{Max Planck Institute for Security and Privacy}
  \city{Bochum}
  \country{Germany}}
\email{steffen.becker@rub.de}

\author{Markus D\"{u}rmuth}
\affiliation{%
  \institution{Leibniz University Hannover}
  \city{Hannover}
  \country{Germany}}
\email{markus.duermuth@itsec.uni-hannover.de}

\renewcommand{\shortauthors}{Kowalewski, et al.}

\begin{abstract}
The COVID-19 pandemic has prompted countries around the world to introduce smartphone apps to \changedcscw{support disease control efforts}. Their \changedcscw{purposes} range from digital contact tracing to quarantine enforcement to vaccination passports, and their effectiveness often depends on widespread adoption. While previous work has identified factors that promote or hinder adoption, it has typically examined data collected at \changedcscw{a single point in time or focused exclusively on digital contact tracing apps}.
In this work, we conduct the first representative study \changedcscw{that examines changes in} people's attitudes towards COVID-19-related smartphone apps \changedcscw{for five different purposes} over the \changedcscw{first 1.5 years of the} pandemic. In three survey rounds \changedcscw{conducted between Summer 2020 and Summer 2021 in the United States and Germany, with approximately 1,000 participants per round and country, we investigate} people's willingness to use such apps, \changedcscw{their perceived utility, and people's attitudes towards them} in different stages of the pandemic. 
\changedcscw{Our results indicate that privacy is a consistent concern for participants, even in a public health crisis, and the collection of identity-related data significantly decreases acceptance of COVID-19 apps.
Trust in authorities is essential to increase confidence in government-backed apps and foster citizens' willingness to contribute to crisis management.
There is a need for continuous communication with app users to emphasize the benefits of health crisis apps both for individuals and society, thus counteracting decreasing willingness to use them and perceived usefulness as the pandemic evolves.}
\end{abstract}

\keywords{COVID-19, privacy, mobile apps, technology acceptance}

\maketitle

\input{sections/1_introduction}
\input{sections/2_relatedwork}
\input{sections/3_method}
\input{sections/4_results}
\input{sections/5_discussion}
\input{sections/7_conclusion}

\clearpage


\appendix





\begin{acks}
This research was supported by Deutsche Forschungsgemeinschaft (DFG, German Research Foundation) under Germany’s Excellence Strategy -- EXC 2092 CASA -- 390781972, by UA Ruhr under the Research Alliance Ruhr program, and by the MKW-NRW Research Training Group SecHuman.
\end{acks}

\bibliographystyle{plain} 
\bibliography{bibliography}

\appendix
\clearpage
\input{tables/deus123_clmm9}
\clearpage
\input{tables/questionnaire_all_waves}
\clearpage

\clearpage
\input{tables/codebooks}

\clearpage
\changedcscw{\section{Willingness to Use Apps and Perceived Utility by App Purpose}}
\label{app:tables-purposes}
\input{tables/deus123_descriptives/Q14_positive_mean.tex}
\input{tables/deus123_descriptives/Q17_utility_mean.tex}

\clearpage
\label{app:interactionplots-purposes}
\input{tables/deus123_descriptives/interactionplots.tex}
\input{tables/deus123_descriptives/interactionplots_concerns.tex}
\input{tables/deus123_descriptives/interactionplots_stategovernment.tex}

\end{document}

%% file: sections/1_introduction.tex
\section{Introduction}
\label{sec:introduction}

Over the course of the pandemic of Coronavirus Disease 2019 (COVID-19), caused by the Severe Acute Respiratory Syndrome Coronavirus 2 (SARS-CoV-2) in early 2020, countries around the world have taken severe measures to contain the spread of the virus. 
These measures include government-mandated quarantines, local and national lockdowns, and contact and travel restrictions, ultimately leading to a breakdown of international travel and the global economy. 
Digital tools were soon developed to support these efforts, with public discussion focusing on smartphone applications. 

In many countries, including Singapore~\cite{sggov_tracetogether_2020}, Israel~\cite{imh_hamagen_2020}, and Germany~\cite{sap_coronawarnapp_2020}, apps were seen as a key factor to relieve the burden on health authorities by assisting their traditional disease management methods with automated digital \emph{contact tracing}.
Apps for other COVID-19-related purposes were deployed by public and private entities worldwide, including apps for \emph{quarantine enforcement}~\cite{hkgov_stayhomesafe_2020}, \emph{symptom checking}~\cite{apple_covidscreeningnews_2020}, and easy access to COVID-19-related \emph{information}~\cite{augov_coronavirusaustralia_2020}.
As the number of people vaccinated against COVID-19 increased and countries sought to lift restrictions and reinstate international travel, the discussion had shifted towards digital \emph{health certificates} that help individuals manage the results of recent tests for SARS-CoV-2 or confirm their vaccination against COVID-19. Examples include the European Union's Digital COVID certificate~\cite{eu_covidcertificate_2023}, which was implemented, for example, into the German CovPass app~\cite{rki_covpass_2021}, the IATA Travel Pass Initiative~\cite{iata_travelpass_2021}, and New York State's Excelsior Pass~\cite{nygov_excelsior_2021}. 
While these apps serve widely different purposes, they all pursue the same common goal: to assist traditional measures in the fight against the COVID-19 pandemic. Our work revolves around such \emph{COVID-19 apps}, which is the umbrella term we use to denote smartphone applications \emph{specifically designed} to help combat the COVID-19 pandemic.

Depending on their concrete functionality, the effectiveness of such apps can rely on widespread voluntary adoption. This was particularly evident with digital contact tracing~\cite{ferretti_contacttracing_2020} but also applies to other COVID-19 apps such as digital health certificates, which can only significantly speed up mandatory checks in everyday life if many people use them. 
The deployment and promotion of such apps by governments has led to  public debates about the privacy, security, and social implications of COVID-19 apps~\cite{panetta_debateitaly_2020, criddle_split_2020, zhang_covidprivacy_2020}, which can influence people's willingness to use them.

\medskip 

While previous work has identified factors that promote or hinder adoption of COVID-19 apps, these studies were typically based on data collected at a single point in time~\cite{altmann_acceptability_2020, bachtiger_belief_2020, li_covidapps_2020, toch_surveillance_2021} or focused solely on contact tracing apps~\cite{simko_contacttracing_2020}. 
Learning which factors promote or hinder adoption of apps for different purposes in the COVID-19 pandemic and how they evolve over time can provide insights for the development of smartphone apps designed to help combat future public (health) crises. 
On a more general scale, these results can inform the development of apps, particularly those issued by non-commercial entities such as governments or NGOs, that collect users' personal information not for their immediate personal benefit but that of society as a whole and, thus, would benefit from widespread adoption. Examples for which such data collection could be beneficial include crisis prediction and management, improvement of local infrastructure, or scientific research.

To better understand how different factors such as app purpose, collected data, technical implications, stance towards authorities, privacy attitudes, prior experience with COVID-19, and demographics influence people's willingness to use such apps, we conduct a sequence of surveys over the course of the first 1.5 years of the COVID-19 pandemic. In three different survey rounds, each administered in both the United States and Germany between June 2020 and May 2021, we use a vignette design of hypothetical yet realistic COVID-19 apps for five different purposes (contact tracing, health certificate, information, quarantine enforcement, and symptom check) to examine the factors that influence people's willingness to use such apps and how these factors change over time, and attempt to relate the results to different pandemic events. More specifically, we investigate the following research questions:

\begin{itemize}

\item[\textbf{RQ\,1}:] \emph{How does people's willingness to use COVID-19 apps for different purposes change over time and what is their perceived utility?} 

We find that for most app purposes there is a slight negative trend in participants' willingness to use COVID-19 apps over time. Across all survey rounds, the perceived utility of apps for each individual purpose is slightly higher than participants' willingness to actually use that app.

\item[\textbf{RQ\,2}:] \emph{What factors significantly influence people's willingness to use COVID-19 apps and (i) are of continuous importance for participants or (ii) whose influence changed during the first 1.5 years of the pandemic?} 

Factors related to the data processing practices of COVID-19 apps tend to have consistently significant influence on adoption. The same applies to apps for contact tracing that were widely discussed and used to make an individual contribution to fight the pandemic.
The impact of other factors on participants' willingness to use apps changed over time, particularly due to individual experience with COVID-19 and external influences, such as the app purpose ``health certificate'' in the light of public discussion of vaccine passports in Germany.

\item[\textbf{RQ\,3}:] \emph{How does people's subjective perception of COVID-19 apps change over time, based on 1.5 years of experience with the pandemic and actual use of these apps?}

Participants' perception of COVID-19 apps has become less extreme over time, with privacy concerns remaining important but being overshadowed by considerations of utility and misconceptions about the capabilities and use cases of apps, including post-vaccination use.

\end{itemize}

Our findings can help better understand how people interact with state-issued technology over the course of a public health crisis with changing external influences, including steeply rising incidence rates and measures for containment that limit individual freedoms. Insights from this can inform the design of future government-issued apps that collect personal information for the benefit of society as a whole in other contexts---such as disaster prevention and management and scientific research---as opposed to providing useful functionality only for the individual. In particular, we identify the importance of establishing trust in the entities behind an app and communication about its capabilities and continuous utility to be vital in achieving widespread voluntary adoption.

%% file: sections/2_relatedwork.tex
\section{Related Work}
\label{sec:related-work}

Our work relates to previous studies of smartphone applications that process personal health information, particularly in the COVID-19 pandemic, and of apps used in crisis situations to inform and communicate with the general public.

\subsection{User Perception of COVID-19 Apps}

Security and privacy research has started to investigate user perceptions of COVID-19 apps as early as spring 2020, initially focusing on apps for digital contact tracing to accompany intense public discussions about different proposed architectures and their implications for security, privacy, and society~\cite{criddle_split_2020}.
In the early phase of the pandemic in April 2020, two studies with international participants found that people were generally highly willing to use such apps, even though security and privacy concerns may be hindering factors~\cite{altmann_acceptability_2020, simko_contacttracing_2020}. 
In a longitudinal follow-up from April to December 2020, Simko et al.~\cite{simko_longitudinal_2022} identified diverse public opinions of contact tracing apps. While the general willingness to adopt such apps continuously increased, it also became apparent that some people will never use them voluntarily, a finding also confirmed in later work~\cite{haring_never_2021}, including our own~\cite{ utz_covidapps_2021}.
In November 2020, Li et al.~\cite{li_contacttracing_2021} identified prosocialness, COVID-19 risk perception, and technology-readiness to be drivers for adoption and confirmed that privacy concerns reduce the use of contact tracing apps.
By contrast, Seberger et al.~\cite{seberger_postcovid_2021} found that people are willing to set back their own privacy expectations when it serves public health as a greater good.

Regarding other drivers that hinder the adoption of digital contact tracing, Toch and Ayalon~\cite{toch_surveillance_2021} found that people are less likely to voluntarily use such apps in environments with mass surveillance measures in place when they have positive attitudes towards such measures.
Beliefs of being immune also reduced the willingness to use contact tracing apps, even among people whose suspected COVID-19 infections were unconfirmed~\cite{bachtiger_belief_2020}.
Fast et al.~\cite{fast_incentivising_2021} explored different types of incentives to increase the adoption rate of contact tracing apps and found monetary incentives to significantly increase app installations. 
Kahnbach et al.~\cite{kahnbach_quality_2021} systematically reviewed 21 European apps for digital contact tracing and recommended to increase in-app user engagement and to combine multiple features (\eg, contact tracing and venue check-in) to foster future adoption of COVID-19 apps.

Beyond contact tracing apps, in our prior work~\cite{utz_covidapps_2021} we studied people's willingness to use different types of COVID-19-related apps in Germany, the United States, and China and found attitudes towards governmental authorities to significantly influence the willingness to use such apps. 
Marhold and Fell~\cite{marhold_electronic_2021} showed an increased importance of unified digital solutions to prove immunization status via digital vaccination certificates.
In this context, another study by some of the authors of this work~\cite{kowalewski_proofofvax_2022} indicated that users' disposition to privacy plays an important role in the willingness to adopt app-based vaccination certificates in Germany.

In health contexts not related to COVID-19, studies have found nuanced privacy attitudes in the sharing of health data depending on the type of data being shared~\cite{dimatteo_mentalhealthapps_2018} and the receivers of shared health-related data~\cite{nicholas_mentalhealth_2019, gorm_workplace_2016}.

\subsection{Use of Crisis Management Apps}
\label{sec:rw-crisis-apps}

In a more general context, previous research investigated mobile apps designed to provide crisis-related information and enable communication between citizens and authorities prior to the COVID-19 pandemic. 

Appleby-Arnold et al.~\cite{applebyarnold_trust_2019} studied the role of trust in authorities in disaster-related crises in Italy and Germany. They found that when using disaster apps, trust between citizens and authorities is generated through perceptions of shared responsibility and tasks. While they consider trust to be situational, the authors explain it as a cultural factor that is subject to constant change in societies.
A study by Dressel~\cite{dressel_riskcultures_2015} complements prior research about the build of trust in cultures and differs between risk cultures of a given society in case of a crisis. These risk cultures have different implications for crisis management and different conclusions need to be drawn when authorities are tasked with a developing crisis: In an \emph{individual-oriented risk culture}, \ie, where trust in authorities is medium to high, crisis management needs to implement the idea that the state is there and well prepared to tackle disasters. If high or even very high trust is placed in authorities (\emph{state-oriented risk cultures}), the state needs to implement the idea that despite responsible crisis management institutions, individual behavior is equally important for handling a crisis in a meaningful way.

In an online study Reuter et al.~\cite{reuter_katwarn_2017} evaluated the use of crisis apps in Europe based on three popular apps focusing on warning functionality. Participants were found to have a positive attitude towards crisis apps, but one of the biggest contentions for users was the need to install more than one app. Users also seem to be interested in being integrated in crisis management and participating as volunteers to report incidents. However, only 16\,\% of the participants used at least one crisis app (as of 2015).
Other work~\cite{kotthaus_mobilewarningapps_2016} related to mobile warning apps in disaster communication found that the messages presented in mobile warning apps for the purpose of crowd control need to be carefully designed and presented to users to consider strong emotions such as fear that often emerge in crisis situations. The authors argue that existing apps in Germany lack quality and timing, which highlights the importance of well-chosen messages to prevent people from ignoring them due to notification fatigue. Further findings show that malfunctions in these apps can lead to a high number of user complaints.
Kaufhold et al.~\cite{kaufhold_crisisapps_2020} conducted a representative user study in Germany in May 2019 to investigate, among others, the functionality and perceived usefulness of mobile crisis apps. Their results indicate that besides emergency and health-related warnings participants value bidirectional communication in mobile crisis apps. Participants also prefer a single comprehensive app over multiple different ones and are resistant to installing more than one crisis app on their smartphones.

\medskip
In this paper, we extend our earlier work~\cite{utz_covidapps_2021} by bringing in the temporal aspect. We examine user perceptions of COVID-19 apps in two different countries (in contrast to \cite{bachtiger_belief_2020, fast_incentivising_2021, haring_never_2021, kowalewski_proofofvax_2022, li_covidapps_2020, seberger_postcovid_2021, simko_contacttracing_2020, simko_longitudinal_2022, toch_surveillance_2021}) using a comprehensive approach:
We compare five different types of COVID-19 apps (contrast \cite{altmann_acceptability_2020, bachtiger_belief_2020, fast_incentivising_2021, haring_never_2021, kowalewski_proofofvax_2022, li_covidapps_2020, seberger_postcovid_2021, simko_contacttracing_2020, simko_longitudinal_2022, toch_surveillance_2021}) based on large representative samples (contrast \cite{fast_incentivising_2021, kahnbach_quality_2021, seberger_postcovid_2021, simko_contacttracing_2020, simko_longitudinal_2022}) collected over a period of nearly one year during different stages of the pandemic (contrast \cite{altmann_acceptability_2020, bachtiger_belief_2020, fast_incentivising_2021, haring_never_2021, kahnbach_quality_2021, kowalewski_proofofvax_2022, li_covidapps_2020, seberger_postcovid_2021, utz_covidapps_2021, toch_surveillance_2021}).
Previous work studied crisis management applications prior to the COVID-19 pandemic~\cite{applebyarnold_trust_2019, dressel_riskcultures_2015, reuter_katwarn_2017, kotthaus_mobilewarningapps_2016, kaufhold_crisisapps_2020}. By contrast, our research examines the perception of (health) related disaster apps and discusses the influence of participants' trust in actors and institutions in a real global crisis based on participants' willingness to use, their opinion of, and actual usage of (governmental) health apps.
Our findings aim to contribute to the development of future state-issued apps designed to benefit society as a whole and, thus, would similarly benefit from widespread adoption.

%% file: sections/3_method.tex
\section{Method}
\label{sec:method}

Our work extends this prior research on COVID-19 apps and crisis management apps by exploring the impact of different app purposes and various factors related or unrelated to the data processing practices of an app on people's willingness to use apps in the context of a global (health) crisis and how their perception and use of such apps developed over the course of the first 1.5 years of the pandemic. In particular, it continues the research from our first paper~\cite{utz_covidapps_2021} by adding in the temporal aspect. Broadly considering people's willingness to use state-supported smartphone apps for crisis management and communication over time, \ie, over different phases of a crisis, can inform the development of similar future apps issued by public actors that collect personal information for the benefit of society as a whole and investigate how external effects beyond app design influence people's acceptance and use of such digital tools.

To this end, we conducted three rounds of online surveys with participants from the United States and Germany, in summer 2020, late fall 2020, and spring 2021, selected to reflect different stages of the COVID-19 pandemic. The data from the first round is the data collected for our first paper~\cite{utz_covidapps_2021} in these two countries, \ie, without the data from China, as we only conducted the first survey round there.

Correspondingly, our survey instrument is the questionnaire we already used in this first survey round and publication and a more detailed description of its creation can be found in that paper~\cite{utz_covidapps_2021}. In its main part it presented participants with vignettes~\cite{finch_vignette_1987} describing (hypothetical) COVID-19 apps and asked them to rate these apps according to a set of criteria. 
Participants were also asked about their smartphone use, experience with the coronavirus, use of COVID-19 apps, privacy concerns, and attitudes towards governmental actions.
In subsequent survey rounds we included additional questions, for example, regarding the official German app for digital contact tracing, the Corona-Warn-App~\cite{sap_coronawarnapp_2020}. However, as the focus of this paper are changes in attitude towards COVID-19 apps over the course of the first 1.5 years of the pandemic, we did not analyze all survey questions. 
In the following, we describe the theoretical framework behind our vignettes and the study protocol: questionnaire, vignette design, recruitment process, methods used for data analysis, and the limitations of our study design.

\subsection{Theoretical Framework: Privacy as Contextual Integrity}
\label{sec:contextual-integrity}

Nissenbaum's theory of privacy as contextual integrity (CI)~\cite{nissenbaum_privacy_2004} has proven to be useful in identifying factors that influence individuals' privacy perceptions~\cite{emaminaeini_iot_2017, benthall_contextual_2017}.
The CI theory states that privacy can be understood as a matter of appropriateness of information flows that is governed by informational and social norms consistent with a number of parameters:
\begin{itemize}
\item Actors (Who will send and receive the data?)
\item Information types (What types of data are concerned?)
\item Transmission principles (What are the means of the data transmission?)
\end{itemize}
Public discussions about anonymity, data recipients, and possible architectures of COVID-19 apps implicitly referred to these parameters.
Like other previous research that studied user willingness to adopt new technology~\cite{apthorpe_iottoys_2019, emaminaeini_iot_2017, martin_records_2017}, we used a vignette design to describe scenarios that vary these parameters and potentially impact participants' willingness to use different variants of COVID-19 apps.

\subsection{Study Protocol}

At the beginning of the survey we informed participants about the purpose of the study, the data collection, and asked for their consent before they proceeded to the actual questionnaire, which consisted of three parts:
\begin{enumerate*}[label=(\roman*)]
    \item a warm-up that asked participants about their smartphone use, experience with COVID-19, and their knowledge and general perception of COVID-19 apps,
    \item the vignette part, where participants saw and rated ten scenarios describing fictitious COVID-19 apps inspired by real ones from all over the globe, and
    \item COVID-19 apps and concerns, where we asked participants about their experience with real-world COVID-19 apps, attitudes towards governmental actions, and general privacy attitudes.
\end{enumerate*}

\subsubsection{Warm-up Questions}

To help participants ease into the questionnaire, its first part asked about their background regarding phone use, COVID-19, and general questions about apps designed to help fight the pandemic.

\paragraph{Smartphone Use}
First, we asked participants about their use of smartphones: if they owned one (Q1), its operating system (Q2), and how happy they were with certain aspects of their device (\eg, its battery life; Q3).

\paragraph{COVID-19 Experience}
Personal experience with COVID-19 was shown to be a significant factor in the adoption of (hypothetical) contact tracing apps~\cite{zhang_covidprivacy_2020}, so we asked about participants' experience with previous infections (Q4 and Q5), quarantine (Q6), and their concern of loved ones getting infected (Q7). We also asked participants about people at higher risk in their household (Q8), and, in the third survey round, vaccination status (Q9).

\paragraph{Knowledge and Perception of COVID-19 Apps}
Next, we asked participants whether they knew any app available in their country for contact tracing, symptom checks, quarantine enforcement, COVID-19 information, or health certificates (Q10), \changedminor{which are the five app purposes of interest in our study}. 
In open-ended questions, we let participants name general positive (Q11) and negative (Q12) aspects of COVID-19 apps. Facing evidence for dissatisfaction with existing apps, in Round~3 we added a new question (Q13) that asked participants what functionality they would wish for in the ``ideal'' COVID-19 app.

\subsubsection{App Scenarios}
\label{sec:appscenario}

The main part of the questionnaire asked participants to rate hypothetical yet realistic COVID-19 apps according to four different criteria. 
Each participant received a unique set of ten vignettes, short texts which each described one (hypothetical) COVID-19 app. 

\paragraph{Vignette Design}

The vignettes shown in our survey contained short textual descriptions of hypothetical COVID-19 apps based on a common text template, whose blanks were filled in to create a concrete app description.  
We denote such a specific app description an \emph{(app) scenario}.

Our previous work~\cite{utz_covidapps_2021} describes in more detail how we created the app scenarios and also provides an illustrative example. 
Each scenario is composed of a fixed text template with gaps for eight data processing \emph{factors}, each of which is filled with one \emph{factor level}. The factors are derived from the theory of contextual integrity (CI) (see \Cref{sec:contextual-integrity}) and the factor levels originate in an analysis of real-world COVID-19 apps conducted in April 2020, which is also described in more detail in our first paper.

The factors and factor levels, whose exact number varies by factor, are as follows:

\begin{enumerate}[noitemsep, nosep, leftmargin=*]
    \item \textbf{Purpose (5 levels):} contact tracing, symptom checking, quarantine enforcement, information, health certificate.
    \item \textbf{Data Collected (16 levels):} encounter data, location data, health or activity data (excluding {COVID-19} infection status), {COVID-19} infection status, all combinations of two or three of them, unspecified data, no data.
    \item \textbf{User Anonymity (3 levels):} whether the app collects personal data that allows for unique identification of the individual, collects only demographic data, or collects only data that cannot be used to uniquely identify the user.
    \item \textbf{Data Receiver (6 levels):} health authorities, law enforcement, research institutions, private companies, the public, none.
    \item \textbf{Data Transmission (3 levels):} automatically, manually (app-type-specific wording, \eg, for health certificate: ``when you request your health report''), none (in case no data is collected).
    \item \textbf{Retention (3 levels):} one month, until end of current coronavirus regulations, unspecified.
    \item \textbf{Technical Implications (3 levels):} impact on battery life, app malfunctioning (app-type-specific wording, \eg, false positive for breaking quarantine), none.
    \item \textbf{Soci(et)al Implications (4 levels):} possible additional benefits in the future, more timely adjustment of local coronavirus regulations, extended personal freedom of movement or travel, none.
\end{enumerate}

\paragraph{Scenario Sets}
Combining all possible factor levels resulted in 155,520 different app scenarios.
As described in more detail in our previous work~\cite{utz_covidapps_2021}, we applied constraints to ensure that the combinations of factor levels made sense, such as symptom check apps always requiring health or activity data or data made available to the public always being stored for an indefinite amount of time.

For each participant, we then created a set of ten scenarios that they would be presented with in the survey in random order. 
To create a set, scenarios were drawn randomly under two constraints:
(i) we included two scenarios for each of the five app purposes, and
(ii) if possible, each pair of scenarios with the same app purpose differed in all other factor levels.

\paragraph{Scenario Questions}

For each app scenario, we asked participants the same four questions. We asked how likely they were to use the described app (Q14) and to estimate the expected share of app users in their country (Q15).
We also let participants assess the normative pressure to use the app (see Ajzen~\cite{ajzen_tbpquestionnaire_2019}) that they expect in their social circles (Q16) and how useful they perceived the described app to be in the fight against the pandemic (Q17).

\subsubsection{COVID-19 Apps \& Privacy Concerns}

\changedcscw{In its third and final part, the survey investigated participants' concrete experience with COVID-19 apps, their attitudes towards public actors and anti-pandemic measures, and towards data privacy on the Internet in general.}

\paragraph{Experience with COVID-19 Apps}
After the app scenarios, we wanted to know whether the participants themselves (had) used such an app (Q18/Q19) and, if yes, which app (Q20/Q22), and if not, why they did not (or no longer) use one (Q21/Q23). We also asked if they were satisfied with the used apps (Q24). German participants who used the national app for contact tracing, the Corona-Warn-App~\cite{sap_coronawarnapp_2020}, were additionally asked how satisfied they were with the app (Q25), if they had ever been warned through it (Q26), and to assess a set of true/false statements about the app (Q27).

\paragraph{Attitudes Towards Governmental Actions}
Since the fight against the pandemic was accompanied by restrictions of personal freedoms in both surveyed countries, we were interested in participants' opinions of and attitudes towards public and governmental institutions.
We let them assess the measures applied in their region to counter the pandemic (Q28) and asked for their opinions of health authorities, law enforcement, research institutions, private companies, and federal and regional governments (Q29).

\paragraph{Individual Privacy Concerns}
Finally, to learn participants' general privacy concerns, we used the Internet Users' Information Privacy Concerns (IUIPC)~\cite{malhotra_iuipc_2004} constructs for Control, Awareness (of privacy practices), and Collection (Q30). 
The IUIPC scores would be used in two contexts: as a demographic factor to describe the sample and as an input factor for regression models.

\subsubsection{Changes to the Questionnaire} 
Between rounds, we aimed to minimize changes to the questionnaire to preserve comparability. To account for the development of the pandemic, we added some questions (such as Q7 about participants' vaccination status) or answer options (such as Luca~\cite{culture4life_luca_2021} for the list of widespread apps possibly used by German participants in Q19).

\subsection{Recruitment}
We \changedcscw{commissioned} an online panel provider, Lightspeed Research (Kantar), to obtain representative samples for the general population of Germany and the US in terms of gender, age, region, and education. 
\changedcscw{
For each survey round, we recruited a new sample of approximately 1,000 participants. We decided against a true longitudinal design with a fixed panel of participants because the duration and further evolution of the pandemic were unpredictable, which would have made it very difficult to plan the overall duration, number, and timing of different survey rounds, \changedminor{as well as the required number of participants under high expected dropout rates between rounds.}

We were still able to compare results across rounds because each sample was representative for the general population with regard to the above criteria in the respective country. 
}
All direct interaction with participants was handled by the panel provider, including setting representative quotas, reimbursement, and data cleaning.
The latter involved discarding respondents who had completed the survey faster than 40\,\% of the median response time or who had responded in suspicious or repetitive patterns. 
\changedminor{The cost was 2,500\,\EUR per survey round and country plus a one-time 2,500\,\EUR for setup and implementation.}

\subsection{Survey Timeline and Pandemic Context}
To provide context for our study, we describe the pandemic situation and availability of COVID-19 apps in the surveyed countries at the time of each survey round.

\begin{figure}[t]
\centering
\includegraphics[width=\columnwidth]{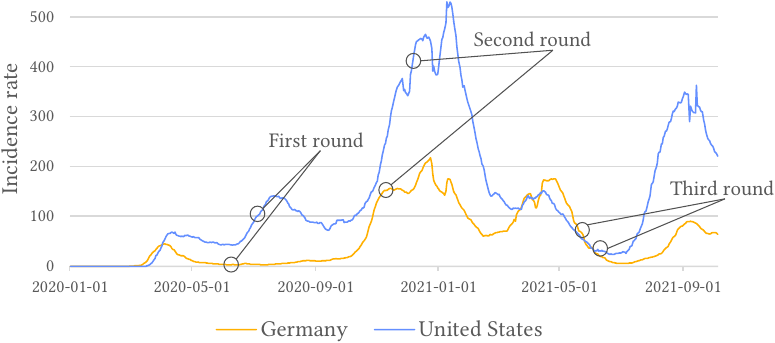}
\caption{Survey timeline for Germany and the US, including COVID-19 incidence rates (new infections per 100,000 people in the last 7 days)~\cite{gcdl_incidencelevel_2023}.} 
\label{fig:incidencelevel}
\end{figure}

\subsubsection{Germany} 
\label{sec:germany}

\changedminor{We started each of the three survey rounds in Germany, as this initially presented us with the opportunity to learn about people's opinions before the release of a federal COVID-19 app.}

\subsubsection*{Round 1 (June 9--11, 2020)} The first \changedminor{round} was conducted during the first wave of the COVID-19 pandemic (see \Cref{fig:incidencelevel}) when strict contact restrictions were in place across the country. This was just days before the launch of the \changedminor{Corona-Warn-App}~\cite{sap_coronawarnapp_2020}, Germany's official smartphone app for digital contact tracing, published on June 16, 2020. Public discussion in the preceding months had been dominated by the privacy risks of a centralized architecture for digital contact tracing~\cite{lomas_centralized_2020}, which ultimately led to the adoption of a decentralized system based on Google's and Apple's Exposure Notification API~\cite{applegoogle_ppct_2020}. At the time of survey Round~1, one COVID-19 app officially recommended by German authorities was available, Corona-Datenspende (``Corona Data Donation'')~\cite{rki_coronadatenspende_2020}, which allowed owners of fitness trackers to voluntarily provide the RKI\footnote{The Robert Koch Institute (RKI) is the leading biomedical research institution of the German Federal Government.} with health and activity data to aid COVID-19 research. Pandemic-related information had also been added to Germany's official disaster alert app, NINA~\cite{bbk_nina_2023}.

\subsubsection*{Round 2 (November 5--18, 2020)} The second round was conducted at the beginning of the second pandemic wave in Germany when incidence rates were steeply rising. At this point in time the \changedminor{Corona-Warn-App} had been available for five months, accompanied by a national advertising campaign. Subsequent updates had fixed bugs and increased usability.

\subsubsection*{Round 3 (May 19--31 2021)} The third survey round took place during the third wave of the pandemic in spring 2021. The national vaccination campaign was ongoing, with 39.0\,\% of the German population being partially and another 12.6\,\% being fully vaccinated~\cite{cdc_coviddatatracker_2023}. New functionality had been added to the \changedminor{Corona-Warn-App}, including a contact diary and management of test results.
In addition to the publicly funded apps, Luca~\cite{culture4life_luca_2021}, an app for digital contact tracing developed by a private company, had been integrated into \changedminor{the anti-pandemic strategies of} multiple German states. This app provided event hosts and gastronomers with a QR-code-based implementation of their legal obligation to record attendance for contact tracing purposes and was directly connected to \changedminor{the computer systems of} health authorities. It has been criticized for its centralized architecture, privacy, and security issues~\cite{stadler_luca_2021}. In addition, public discussion at the time of Round~3 revolved around app-based digital vaccine certificates to let users easily prove their COVID-19 vaccination status~\cite{cpc_debatevaxcert_2020}. This functionality was ultimately integrated into the CWA on June 9 and issued as a standalone app, CovPass~\cite{rki_covpass_2021}, on June 10.

\subsubsection{United States} 
\label{sec:usa}

After Round~1 in the US had been conducted approximately two weeks after Round~1 in Germany, we followed a similar schedule for subsequent rounds.

\subsubsection*{Round 1 (July 6--14, 2020)}
In the US, Round~1 took place during the first wave of the COVID-19 pandemic. High incidence rates had prompted multiple states including California and Indiana to postpone or reverse plans to reopen their economies~\cite{cdc_coviddatatracker_2023}.
A coronavirus-related app available in the US at that time was Apple's now discontinued COVID-19 Screening Tool~\cite{apple_covidscreeningnews_2020} that provided information about the disease and allowed users to assess their symptoms.

\subsubsection*{Round 2 (December 7--17, 2020)}
The second round was conducted during the second pandemic wave with an exponential increase of infection rates. Starting in November 2020, this development had led several states, such as Michigan, Washington, and California, to impose strict measures including the closure of high schools and restaurants~\cite{bbc_usrestrictions_2020}.
Unlike in Germany, there had not been any plans in the US to roll out a contact tracing app on the federal level. Instead, from August 2020 US states, territories, and Washington, D.C. started to issue their own apps, some with a common code base and the ability to interoperate~\cite{nyczepir_ctapps_2021, sato_contacttracing_2021}. Examples included Virginia's COVIDWISE app, published on August 5, and California's CA Notify~\cite{cadoh_canotify_2023}, available since December 10. 
December 2020 also marked the start of mass vaccinations against COVID-19 in the US.

\subsubsection*{Round 3 (June 7--23, 2021)}
The final survey round was conducted at a time when
the vaccination campaign in the US had reached about half of the total population, with approximately 50.9\,\% having received at least one dose of the vaccine and 41.5\,\% being fully vaccinated~\cite{cdc_coviddatatracker_2023}.
Several US states had issued bans against basing access to certain goods and services on COVID-19 vaccination status~\cite{davis_vaccinepassports_2021}. Only the state of New York had issued a vaccine passport app as of US Round~3, Excelsior Pass (Plus)~\cite{nygov_excelsior_2021}.
Additionally, private entities had started to issue health certificate apps, such as the IATA Travel Pass published on April 15, 2021 (and decommissioned by the time of writing), which could be used to provide proof of vaccination for air travel and was recognized by 290 different airlines~\cite{iata_travelpass_2021}.

\subsection{Data Analysis}

Our data analysis process comprised quantitative and qualitative methods.
We focus on changes over time and only briefly describe differences between countries, which we already analyzed in our previous work~\cite{utz_covidapps_2021}.

\subsubsection{Statistical Analysis}
In our evaluation we first investigated participants' overall willingness to use COVID-19 apps. To examine differences in the willingness to use apps (Q14) across three rounds and two countries, we used analyses of variance (ANOVA).
In order to understand the effects of different app scenario factors on participants' willingness to use an app, we performed a regression analysis with the cumulative link mixed models (CLMM) module of the R package \texttt{ordinal}~\cite{christensen_cumulative_2018}. We were specifically interested in how effect sizes change over time, so we included interactions between several factors and the survey round, with Round 1 as the baseline. Since a cross-combination of all factors would have resulted in a model that is not computable, we first conducted a manual visual review of the interactions between each factor and the answers to Q14 to determine which factors yielded interesting results. The resulting models are quite large and included 43 individual factors and 27 (Germany) / 29 (US) interactions, respectively. We followed best practice~\cite{zuur_mixed_2009} to improve the model and iteratively removed non-significant factors, as indicated by the removal increasing the Akaike information criterion (AIC) of the model compared to a version that included that factor or interaction. 

\subsubsection{Qualitative Analysis}
\label{sec:method-qualitative}

Our survey contained seven open-ended questions (Q11--Q13 and Q20--Q23, see \Cref{appendix:questionnaire}). For analysis, we applied Mayring's mixed-methods approach with qualitative and quantitative elements~\cite{mayring_qualitative_2014}: We first used an iterative open coding procedure to create a codebook for each question and subsequently applied it to the data. In the second step, we determined and compared code frequencies across countries and survey rounds.

The coding procedure was performed by two of the authors. First, each coder independently examined the first 300\footnote{In Round~3, only the first 200 answers were used to create the new codebook for Q13 (ideal COVID-19 app).} answers to each question to inductively identify and categorize recurring themes and create an initial set of codes. These were compared and discussed to create a first codebook for each question. Each answer could be assigned one or multiple codes. 

To validate the codebooks, the next 200 answers to each question were independently coded by both researchers. This accounts for roughly 20\,\% of the data, which falls within the recommended amount of 10--25\,\% to determine inter-coder reliability (ICR)~\cite{oconnor_icr_2020}. As a metric for this we used ReCal 2~\cite{freelon_recal_2010} to compute Krippendorff's alpha for each code and, for each codebook, report their mean, weighted by how often the respective code occurred in the ICR data. 

The remaining 500--600 answers were then split into two and each coded by a single coder.

In subsequent survey rounds, we used the first 200 answers to each question to determine if the previous codebooks still applied or needed additions. In Round~3, the ongoing COVID-19 vaccination campaign led us to add ``vaccination'' as a reason named by participants to not (or no longer) use an app. Otherwise, we applied the same coding procedure as in Rounds~1 and~2.

For open-ended responses to Q20/Q22 that asked which apps participants used, we used a simplified process: Using Internet searches and app stores to identify any app previously unknown to us, we mapped each answer to one of the following categories: COVID-19 app, other health app, disaster app, other non-health app, unnamed app (\eg, when participants could not remember the name of an app), a non-app tool (such as a website), none (when participants contradicted themselves and wrote into the text field that they did not use any app or tool), or unusable (incomprehensible).

\subsection{Research Ethics}
Our department does not have an institutional review board. 
Still, we followed best practices of human subject research and data protection guidelines. 
The study was approved by our institution's data protection officer. 
We complied with the rules of the EU General Data Protection Regulation (GDPR) and obtained participants' informed consent at the beginning of the study. 
Our panel provider, Kantar, follows a self-commitment to the ICC/ESOMAR International Code on Market and Social Research~\cite{eccesomar_internationalcode_2023}.

\subsection{Limitations}
\label{sec:limitations}
The various factors and factor levels that could occur in a COVID-19 app were created based on a set of apps from the beginning of the pandemic. Due to the study investigating developments over time, we could not make major changes to the questionnaire or app scenarios to reflect more recent events. In particular, the factors for the CLMM model had to stay the same to allow for an analysis of changes over time.  Thus, we tried to account for possible future developments from the beginning. Nevertheless, the app scenarios presented in the vignettes are theoretical and simplified and thus unlikely to accurately represent existing apps.

Still, in our evaluations we tried to account for changing circumstances over the course of the pandemic, such as social restrictions, newly released apps, and ongoing public discussions about COVID-19 apps. For this, we included questions about measures taken to fight the pandemic and participants’ opinion about state institutions. However, we cannot assign cause and effect beyond doubt for external factors not included in our study, such as political changes like the 2020 presidential election in the US. We did not collect data on political opinions in our questionnaire, because comparing political views across countries is not trivial and this was also not the focus of this work. As we could only speculate about the effect of these external factors, we refrain from discussing them.

%% file: sections/4_results.tex
\section{Results}
\label{sec:results}

Our results are based on the data collected in a total of six different survey rounds, conducted in Germany and the US at three distinct points in time during the first 1.5 years of the COVID-19 pandemic.

\subsection{Participants}
\label{sec:participants}

Over three survey rounds in two countries, we recruited a total of 6,124 participants, 3,049 from Germany (DE) and 3,075 from the US, in six independent samples, each with roughly 1,000 participants as shown in \Cref{tab:demographics}. 
Target representativity quotas were met with average discrepancies of $ 2.5\,\% $ (DE, Round~1), $ 1.0\,\% $ (DE, R2), $ 2.4\,\% $ (DE, R3), $ 3.4\,\% $ (US, R1), $ 4.5\,\% $ (US, R2), and $ 4.2\,\% $ (US, R3).

\input{tables/all123_participants}

\subsubsection{Demographics}

\Cref{tab:demographics} further shows the demographics of our participants (gender, age, and education as reported by our panel provider), the percentage of participants who used at least one COVID-19 related app, and the mean values of their IUIPC scores.

Across all rounds and countries, IUIPC scores were above the median of the {7-point} Likert scale (median = 4). The scores for the three IUIPC dimensions were similar for all rounds in the US, with scores lowest for Collection. In Germany, IUIPC scores decreased from Round~1 to Round~3 for all dimensions, and scores were lowest for Awareness.

\subsubsection{COVID-19 Apps Used by Participants}
\label{sec:whichapp}

\Cref{tab:demographics} shows that the number of German participants who reported use of COVID-19 apps (3.99\,\% in Round~1) almost increased tenfold with the availability of the Corona-Warn-App in Round~2 (38.9\,\%) and continued to grow in Round~3 (42.7\,\%). In the US, usage first increased between Round~1 (6.2\,\%) and Round~2 (11.3\,\%), but then dropped between Rounds~2 and 3 (8.6\,\%).
One possible reason could be the quickly progressing COVID-19 vaccination campaign in the US, as our qualitative analysis in \Cref{sec:whynoapp} and~\ref{sec:whydeleted} hinted at some people thinking an app was no longer necessary if one had received the vaccine.

In Germany in Round~1 (R1), 40 participants reported use of a COVID-19 app, but when asked which, only 11 named a dedicated COVID-19 app, most frequently (7 times) Corona-Datenspende. In R2, where 396 participants (38.90\,\%) had indicated to use any kind of COVID-19 app in the multiple-choice Q19, 325 reported use of the Corona-Warn-App (CWA), 44 of Corona-Datenspende, and 111 of other apps, including contact tracing apps of other countries when on vacation. 64 participants indicated to have deleted a previously used app (50 CWA, 13 Datenspende, and 7 another app).  In R3, 439 participants (42.70\,\%) used a COVID-19 app, broken down into concrete apps as follows: 369 CWA, 70 Datenspende, 172 Luca, 82 other, including vaccine passports and apps showing current COVID-19 regulations. 107 participants had deleted an app (72 CWA, 22 Datenspende, 24 Luca, and 17 another app).

In the US, 62 participants (6.18\,\%) indicated to use a COVID-19 app (Q18) in R1, but only 7 out of 47 open-ended answers to Q20 named a COVID-specific app, most frequently (3 times) the app How We Feel. R2 yielded 113 (11.30\,\%) self-reported app users, and out of 96 open-ended answers 30 used designated COVID-19 apps, most frequently state-specific contact tracing apps like Covid Alert NY, CA Notify, or Slow Covid NC.
For R3 we counted 92 (8.58\,\%) affirmative answers to Q18 and as many open-ended responses, with 19 answers referring to COVID-specific apps, again most being state-specific apps.
The majority were contact tracing apps, but one participant named New York State's vaccine passport app.

\subsection{Willingness to Use and Perceived Utility of COVID-19 Apps Over Time (RQ\,1)}
\label{sec:w-to-use-rq1}

The first part of our evaluation focuses on participants' general willingness to use different types of COVID-19 apps over time.

\paragraph{General Willingness to Use Apps Over Time} First we take a look at how willing participants were across survey rounds to use any kind of COVID-19 app, irrespective of purpose. 
\Cref{tab:willingness-to-use-mean} shows the average response values on the 7-point Likert scales in Q14 (willingness to use the presented hypothetical COVID-19 apps) along with standard deviations across all app scenarios by survey rounds and countries.

Overall, participants' willingness to use an app is rather low in both countries across all three survey rounds---average responses are always below the neutral response, which is 4 on a 7-point Likert scale. Values change over the course of the pandemic in Germany and the US. 
For Germany, we found pairwise statistically significant differences ($ p < 0.05 $) between R1 and R2, as well as between R2 and R3. 
This shows that in Germany the willingness to use COVID apps decreased significantly from R1 to R2 and increased significantly again from R2 to R3.
There is no significant difference in the willingness to use any app between R1 and R3. 
In the US, the willingness to use any COVID-19 app dropped significantly in R3 compared to both other rounds.
These differences between R1 and R3 as well as between R2 and R3 were significant ($p<0.05$).
We did not observe statistically significant differences between R1 and R2.
Pairwise comparisons of the average willingness between both countries for each round (\ie, R1-DE vs. R1-US), always yield significant differences between German and US participants.

\begin{table}[h]
\caption{Mean response values ($\pm$ sd) related to the general willingness to use COVID-19 apps (Q14). Data marked with an asterisk is significantly lower ($ p < 0.05 $) than in the two other rounds in the same country.}
    \centering
    \begin{tabular}{lccc}
    \toprule
    & \textbf{Round 1} & \textbf{Round 2} & \textbf{Round 3} \\
    \midrule
         \textbf{Germany} &  $3.25 \pm 2.04$ & $2.94 \pm 2.01^{\ast}$ & $3.24 \pm 2.04$ \\
         \textbf{United States} & $3.12 \pm 2.10$ & $3.14 \pm 2.16$ & $3.03 \pm 2.16^{\ast}$ \\
    \bottomrule
    \end{tabular}
    
    \label{tab:willingness-to-use-mean}
\end{table}

\paragraph{COVID-19 Apps by Purpose} Next, we more closely examined the different types of COVID-19 apps and evaluated participants' willingness to use these apps (Q14) and their perceived utility (Q17) for each distinct purpose: contact tracing, symptom check, quarantine enforcement, health certificate, and information.
In this analysis we focused on the percentage of positive responses to the two questions, \ie, all responses higher than the neutral response (> 4) on the 7-point Likert scales.
\Cref{tab:Q14Q17_positive} shows these percentages by country, survey round, and app purpose.
Corresponding mean response values to the two questions can be found in \Cref{tab:Q9_mean,tab:Q12_mean} in \Cref{app:tables-purposes}.

Across both countries and all three survey rounds, participants' willingness to use (Q14) a contact tracing app is higher than for any other app purpose. 
We speculate that this is an effect of apps for digital contact tracing being prominently discussed and subsequently launched in many countries across the globe. 
In Germany, the willingness to use an app of any type decreases between R1 and R2. The subsequent increase in R3 is less pronounced, resulting in an overall negative trend across all survey rounds.
In the US, there is a continuous negative trend over all rounds. 
We suspect that the temporary decrease in German participants' willingness to use apps across all purposes in R2 is related to lockdowns that were in place at that time due to high incidence rates (see \Cref{fig:incidencelevel}). 
In this situation, participants could have considered COVID-19 apps as being unable to make a distinct contribution in containing the pandemic and not offering societal or individual benefits~\cite{bosen_lockdown_2021}. 
Similarly, the subsequent increase in R3 could reflect the turn in public policy at that time--released lockdowns and lower incidence rates prompting people to socialize more frequently again.
While this development can be observed for all types of apps, it is especially prominent for health certificate apps in Germany, where the willingness to use this type of app in R3 (\SI{31}{\percent}) was even higher than in R1 (\SI{26}{\percent}). 
A possible reason for this could be that German participants associated this app purpose, initially inspired by the Health Code systems deployed in multiple Chinese municipalities~\cite{utz_covidapps_2021}, with vaccination and test certificates. At the time of survey R3, there were public discussions about the introduction of (digital) ``vaccine passports'' and whether they should be made mandatory for various aspects of public life in Germany~\cite{jordans_privileges_2021, oltermann_jab_2021, wilf-miron_incentivizing_2021}.

\begin{table}[tb]
\definecolor{tabcol}{RGB}{80,140,200}
\caption{Percentage of positive responses to Q14 (willingness to use COVID-19 apps) and Q17 (perceived utility of COVID-19 apps).}
\centering
\begin{tabular}{lccc@{\hspace{0.4cm}}ccc@{\hspace{0.8cm}}ccc@{\hspace{0.4cm}}ccc}
\toprule
\textbf{App Purpose} & \multicolumn{6}{c}{\textbf{Willingness to Use App}} & \multicolumn{6}{c}{\textbf{Perceived Utility}} \\
\cmidrule{2-13}
&\multicolumn{3}{l}{Germany} & \multicolumn{3}{l}{United States} &\multicolumn{3}{l}{Germany} & \multicolumn{3}{l}{United States} \\
\cmidrule{2-13}
& R1 & R2 & R3 & R1 & R2 & R3 & R1 & R2 & R3 & R1 & R2 & R3 \\
& \% & \% & \% & \% & \% & \% & \% & \% & \% & \% & \% & \%\\
\midrule
Contact Tracing & \cellcolor{tabcol!74} $37$ & \cellcolor{tabcol!62} $31$ & \cellcolor{tabcol!68} $34$ & \cellcolor{tabcol!64} $32$ & \cellcolor{tabcol!64} $32$ & \cellcolor{tabcol!58} $29$ & \cellcolor{tabcol!82} $41$ & \cellcolor{tabcol!70} $35$ & \cellcolor{tabcol!74} $37$ & \cellcolor{tabcol!78} $39$ & \cellcolor{tabcol!74} $37$ & \cellcolor{tabcol!64} $32$ \\  
Health Certificate & \cellcolor{tabcol!52} $26$ & \cellcolor{tabcol!42} $21$ & \cellcolor{tabcol!62} $31$ & \cellcolor{tabcol!54} $27$ & \cellcolor{tabcol!54} $27$ & \cellcolor{tabcol!52} $26$ & \cellcolor{tabcol!62} $31$ & \cellcolor{tabcol!48} $24$ & \cellcolor{tabcol!64} $32$ & \cellcolor{tabcol!62} $31$ & \cellcolor{tabcol!60} $30$ & \cellcolor{tabcol!54} $27$\\ 
Information & \cellcolor{tabcol!60} $30$ & \cellcolor{tabcol!46} $23$ & \cellcolor{tabcol!54} $27$ & \cellcolor{tabcol!52} $26$ & \cellcolor{tabcol!56} $28$ & \cellcolor{tabcol!50} $25$ & \cellcolor{tabcol!66} $33$ & \cellcolor{tabcol!50} $25$ & \cellcolor{tabcol!54} $27$ & \cellcolor{tabcol!58} $29$ & \cellcolor{tabcol!60} $30$ & \cellcolor{tabcol!52} $26$\\
Quarantine Enf. & \cellcolor{tabcol!54} $27$ & \cellcolor{tabcol!40} $20$ & \cellcolor{tabcol!52} $26$ & \cellcolor{tabcol!50} $25$ & \cellcolor{tabcol!52} $26$ & \cellcolor{tabcol!48} $24$ & \cellcolor{tabcol!66} $33$ & \cellcolor{tabcol!52} $26$ & \cellcolor{tabcol!64} $32$ & \cellcolor{tabcol!66} $33$ & \cellcolor{tabcol!62} $31$ & \cellcolor{tabcol!54} $27$\\ 
Symptom Check & \cellcolor{tabcol!64} $32$ & \cellcolor{tabcol!46} $23$ & \cellcolor{tabcol!56} $28$ & \cellcolor{tabcol!58} $29$ & \cellcolor{tabcol!62} $31$ & \cellcolor{tabcol!54} $27$ & \cellcolor{tabcol!70} $35$ & \cellcolor{tabcol!52} $26$ & \cellcolor{tabcol!60} $30$ & \cellcolor{tabcol!68} $34$ & \cellcolor{tabcol!66} $33$ & \cellcolor{tabcol!56} $28$\\ 
\bottomrule
\end{tabular}
\label{tab:Q14Q17_positive}
\end{table}

\begin{figure}[tb]
\centering
    \begin{subfigure}[b]{0.49\columnwidth}
        \includegraphics[width=\textwidth]{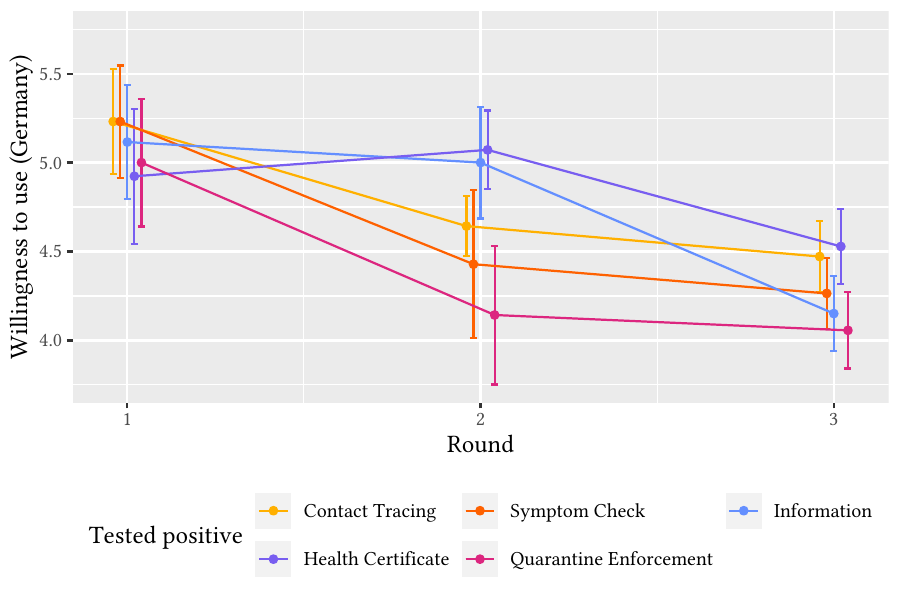}
    \end{subfigure}
    \hfill
        \begin{subfigure}[b]{0.49\columnwidth}
        \includegraphics[width=\textwidth]{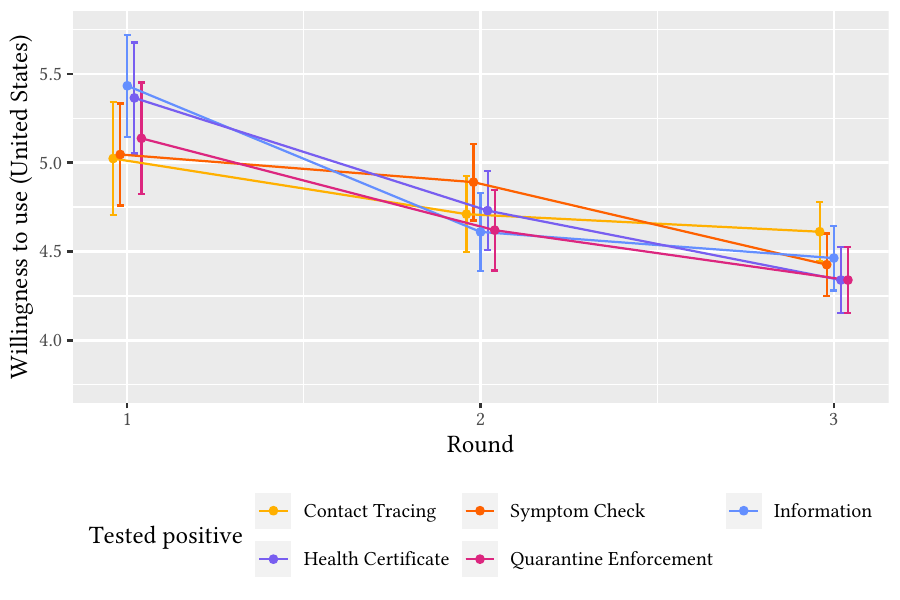}
    \end{subfigure}
        \begin{subfigure}[b]{0.49\columnwidth}
        \includegraphics[width=\textwidth]{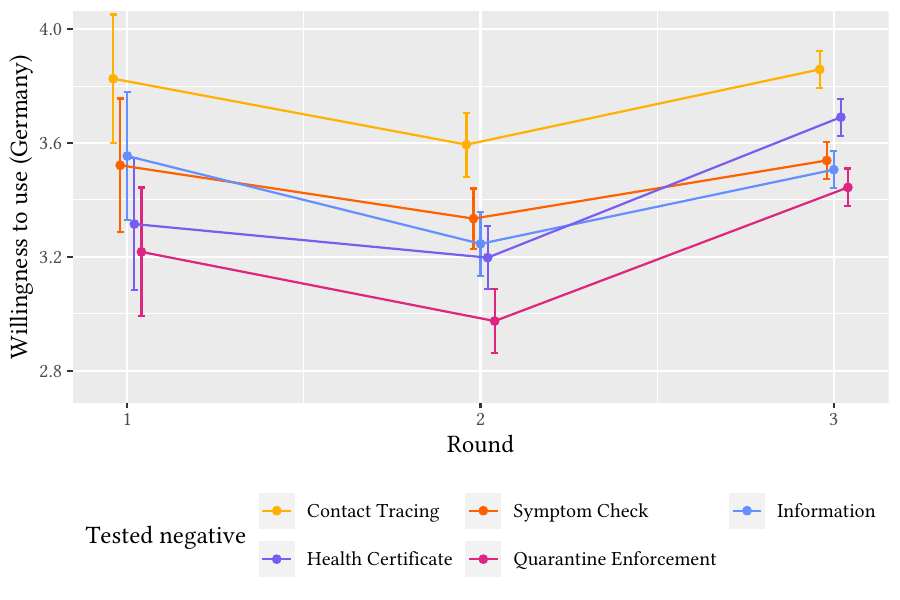}
    \end{subfigure}
    \hfill
        \begin{subfigure}[b]{0.49\columnwidth}
        \includegraphics[width=\textwidth]{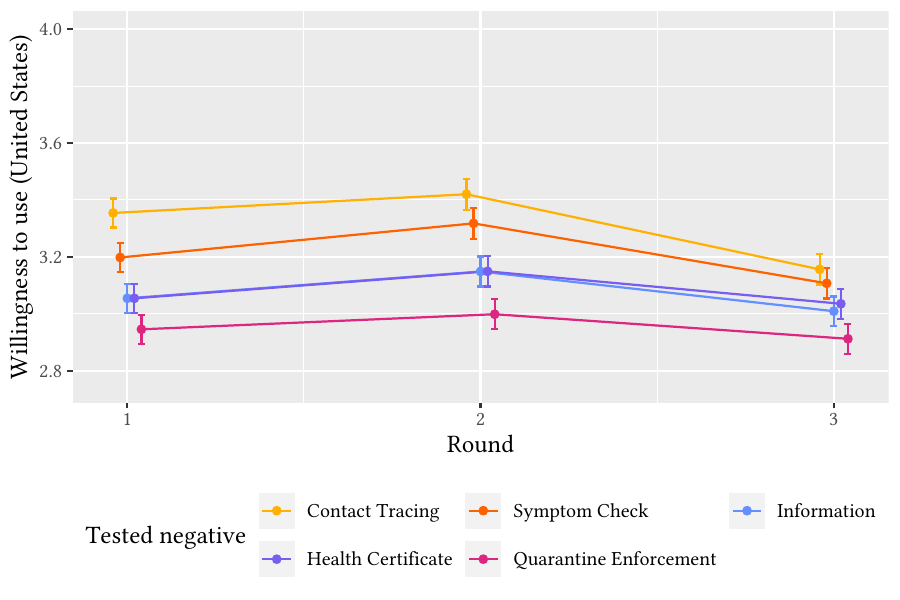}
    \end{subfigure}
    \vskip\baselineskip
    \caption{Willingness to use COVID-19 apps, differentiated by participants who had tested positive (top) or negative (bottom) in Germany (left) and the United States (right).}
    \label{fig:interaction_testedpositivenegative_usde}
\end{figure}

\medskip
In \Cref{fig:interaction_testedpositivenegative_usde} we investigate how participants' COVID-19 infection status affected their willingness to use different types of COVID-19 apps over time. We observe noticeable differences between participants who had reportedly tested positive and those who stated that they had tested negative: In both countries participants who had already been confirmed to have the coronavirus exhibited a higher willingness to use COVID-19-related apps (top figures). Among participants who had not yet contracted COVID-19 (bottom figures), the willingness to use apps is noticeably lower. What stands out is that for negatively tested participants in Germany, the willingness to use contact tracing apps is higher than for the other app purposes, while for participants with a negative test result in the United States the willingness to use both contact tracing and symptom check apps is higher (both bottom figures) compared to those participants who reported having tested positive (top figures). We hypothesize that for individuals who had tested positive, the motivation to further use contact tracing apps in particular was lower than for those who had so far only tested negative and were still trying to avoid an infection. This supports our hypothesis that the perceived individual benefits of COVID-19-related apps influence the decision-making process whether to use an app or not.
Overall, our results indicate that participants who had already had a negative experience with the coronavirus (\ie, had been infected) or who had higher concerns of becoming infected (see \Cref{fig:interaction_concerns_usde} in \Cref{sec:interactionplots}) tended to be more willing to use COVID-19 apps across all survey rounds.

For the perceived utility of COVID-19 apps (Q17), we make similar observations (see the right part of \Cref{tab:Q14Q17_positive}). 
Interestingly, the percentages of participants with positive responses for utility are slightly higher than for the willingness to use the respective type of app. 
This indicates that there is a certain share of participants who consider these apps useful in general but are not actually willing to use the respective app.

\textbf{Summary RQ\,1:} Participants' willingness to use COVID-19 apps is generally rather low, with the average being below the neutral response. For contact tracing, it is higher than for any other purpose in both Germany and the US. Over time, there is a slight negative trend in the willingness to use apps for all purposes, with the exception of health certificate apps in Germany. People with higher concern of contracting the coronavirus are more likely to use apps for digital contact tracing. Participants' perceived utility of COVID-19 apps is slightly higher than their willingness to use them.

\subsection{Factors Influencing the Willingness to Use COVID-19 Apps Over Time (RQ\,2)}
\label{sec:clmmmodel}

As described in \Cref{sec:method}, we used ordinal regression analysis to measure the impact of each app scenario factor on participants' willingness to use COVID-19 apps and how this influence evolved over time.
In this analysis we distinguish between \emph{scenario-specific factors} that originate in the ten app scenarios shown to participants and \emph{non-scenario factors}, which are the responses to the survey questions outside the app scenarios plus demographic information provided by Kantar. In presenting the influence of factors we further differentiate how significant this influence is over time:

\begin{itemize}
    \item factors that significantly influence the willingness to use apps across all survey rounds, \ie,
    that are of continuous importance to participants (highlighted \factor{red} in our models), and 
    \item factors that have significant influence in one or multiple survey round(s) but not in others, \ie, that change in interaction with the survey round (highlighted \factorrounds{blue} in our models).
\end{itemize}

\Cref{tab:clmm_shortened} presents selected significant factors from the two final CLMM models, one for each country, and
\Cref{tab:clmm_appendix} in \Cref{sec:clmm_appendix} shows the comprehensive models with all factors that were included in the respective model.

\subsubsection{Willingness to Use Apps Over Time}

As shown in the first block of rows in \Cref{tab:clmm_shortened}, the influence of time, \ie, the factor ``Survey Round,'' on participants’ willingness to use any of the presented COVID-19-related smartphone apps changed over the course of the pandemic in both Germany and the US, though in a different way: Among German participants, time had a significant positive effect on willingness to use any COVID-19 app (R2: estimates 0.51, $ p < 0.05 $, R3: estimates 0.95, $ p < 0.01 $, compared to R1 as baseline).
In the US, time had a significant negative effect on the willingness to use an app in both R2 (estimates -1.41, $ p < 0.15 $) and R3 (estimates -0.86, $ p < 0.05 $) compared to R1.

Since these results may seem counter-intuitive when compared to the overall willingness to use apps (see~\Cref{sec:w-to-use-rq1}), we emphasize that both measure different effects.
The estimates in \Cref{tab:clmm_shortened} present the actual effect of time on willingness to use an app (which is positive in Germany), whereas the overall results in \Cref{tab:willingness-to-use-mean} show cumulative effects of all factors, resulting in a significantly smaller willingness to use apps in R2 in Germany when compared to R1.

\subsubsection{Influence of Factors on the Willingness to Use COVID-19 Apps Over Time in the US}

As already mentioned above, when investigating the influence of different scenario-based and other factors on US participants' willingness to use COVID-19 apps over time, we differentiate between factors with continuous significant influence on app adoption and those that were not always significant.

\paragraph{Consistently Influential Factors}
Considering only app scenario factors that had consistently significant influence in every survey round  (\factor{red}), we observe in \Cref{tab:clmm_shortened} that contact tracing was the only app purpose with a positive influence on participants' willingness to use an app (Q14) in the US (estimate 0.21, $ p < 0.05 $). Also consistently significant but with a negative estimate were apps reducing phone battery life (estimate -0.12, $ p < 0.05 $) and both identification data levels: demographic data (estimate -0.18, $ p < 0.05 $) and data that allowed for unique identification of an individual (estimate -0.25, $ p < 0.01 $). Participants being worried about sharing their data with private companies and law enforcement were also less likely to use any COVID-19 related app (estimates -0.18, $ p < 0.05 $ and -0.25, $ p < 0.01 $, respectively).

\input{tables/deus123_clmm_shortened_v2.tex}

As shown by the black entries in the full US model in \Cref{tab:clmm_appendix} in \Cref{sec:clmm_appendix}, most app scenario factors (\eg, purpose, data transmission, or retention) did not have any significant influence on US participants' willingness to use an app in interaction with the survey round (\ie, over time). For the factors highlighted red (\eg, whether data can be used to identify an individual, data receiver) we found significant influence on the willingness to use an app in every survey round, which means that these are factors that are of continuous importance for US participants' decision whether to use an app or not. 

\paragraph{Factors with changing influence over time}

Further, our analysis identified factors which we found to not always have significant influence on US participants' willingness to use COVID-19 apps in every survey round (\factorrounds{blue}). 
Regarding app scenario factors, we found such varying influence only for the payload data type ``location'', which had a significant effect in interaction with both survey rounds R2 and R3 (estimate 0.14 and 0.15, $ p < 0.05 $) on the willingness to use an app. This significant influence in R2 and R3 could not be observed in R1 and, thus, constitutes a change in significance over time.

Non-scenario factors (\ie, those taken from parts of the survey outside the app scenarios) more frequently had significant influence on participants' willingness to use an app in interaction with the survey round. One example is the reported education level: In R2 and R3 participants with a high school or higher education were more likely to use any app compared to those with secondary school or less in R1. This effect was higher for participants with a higher formal education level and also higher in R3 than in R2, as indicated by the higher estimates.

\begin{figure}[tb]
\centering
    \begin{subfigure}[b]{0.49\columnwidth}
        \includegraphics[width=\textwidth]{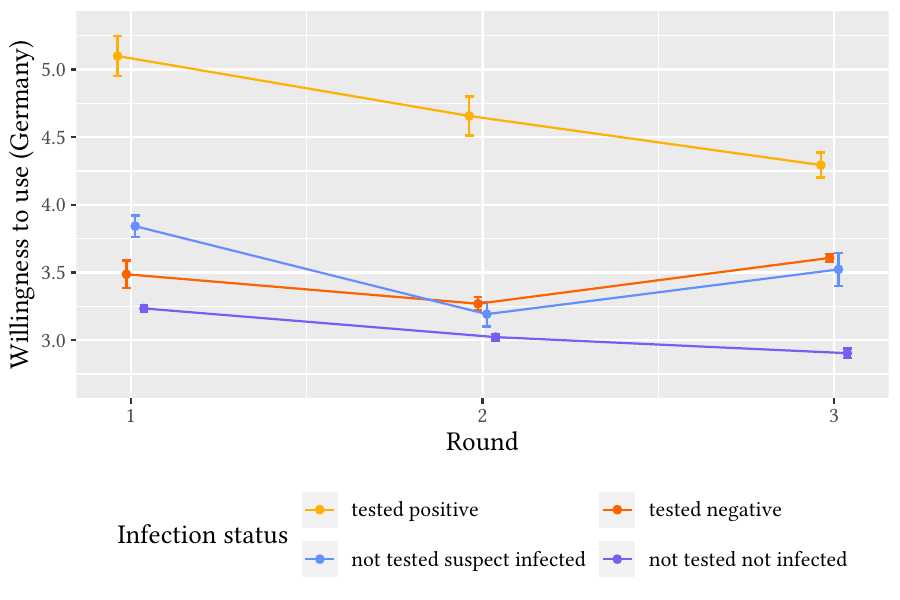}
    \end{subfigure}
        \begin{subfigure}[b]{0.49\columnwidth}
        \includegraphics[width=\textwidth]{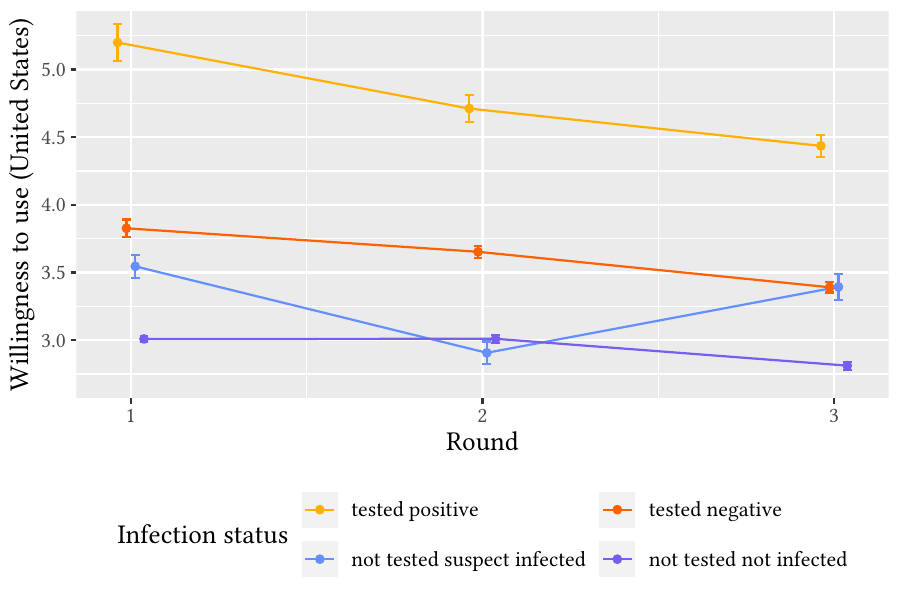}
    \end{subfigure}
    \caption{Willingness to use any app in Germany (left) and the United States (right) in relation to the reported COVID-19 status over time.}
    \label{fig:interaction_infection_usde}
\end{figure}

Another influential factor in interaction with time was the participant's own experience with COVID-19 (Q4), which has an effect in interaction with R3. Here participants who had tested positive were less likely to use any app (estimate: -0.77, $ p < 0.001 $), and those who had not yet been tested but suspected they had already been infected were more likely to use any app (estimate: 0.34, $ p < 0.05 $). \Cref{fig:interaction_infection_usde} shows on the right-hand side the (simplified) dependency between US participants' reported infection status and their willingness to use an app over time.

The analysis over time also shows that in R3 participants of higher age exhibited significantly less interest in using any app, although the effect size is rather small (estimate -0.01; $ p < 0.001 $). In R3 those with an unfavorable opinion of healthauthorities (estimate -0.77, $ p < 0.001 $) or the federal government (estimate: -0.40, $ p < 0.001 $) were less likely to use any app, as did those with a favorable view of their state government (estimate: -0.39, $ p < 0.001 $).

The influence of US participants' privacy concerns as indicated by their IUIPC score also changed significantly over time. Both IUIPC scores of the dimensions Collection and Awareness had a significant influence on participants' willingness to use a COVID-19 app: an increasing willingness to use an app for Collection and a significant decreasing tendency to use a COVID-19 app for Awareness, in both survey rounds R2 and R3 compared to R1 (see~\Cref{tab:clmm_shortened}). 
In the US the willingness to use any variant of COVID-19 app related to the IUIPC dimension Collection significantly decreased over all survey rounds.

\begin{figure}[tb]
\centering
    \begin{subfigure}[b]{0.49\columnwidth}
        \includegraphics[width=\textwidth]{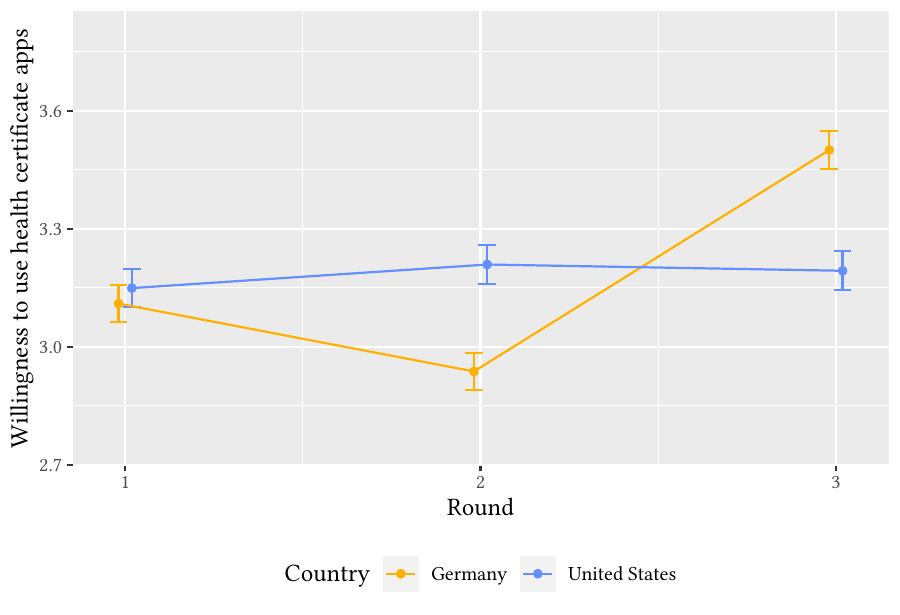}
    \end{subfigure}
        \begin{subfigure}[b]{0.49\columnwidth}
        \includegraphics[width=\textwidth]{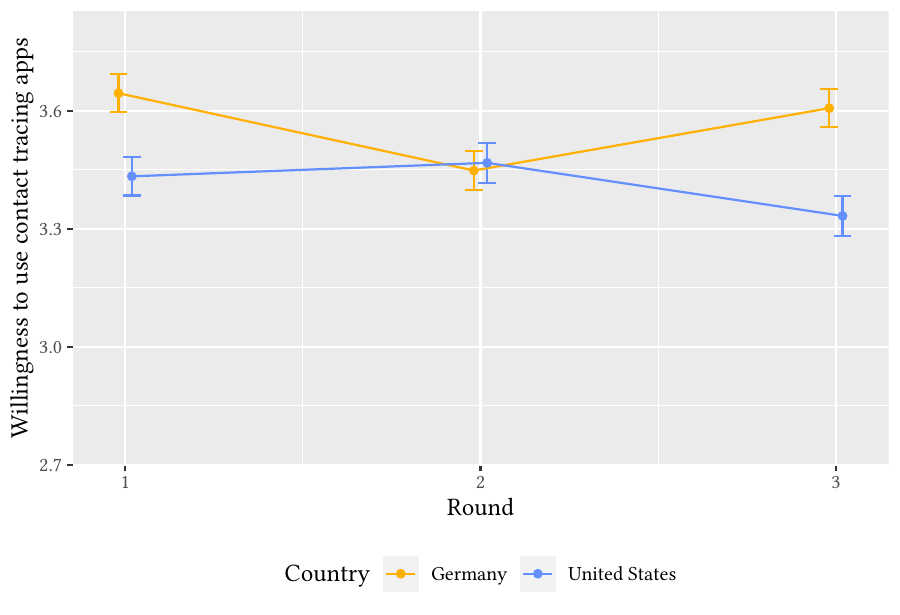}
    \end{subfigure}
    \caption{Willingness to use health certificate (left) and contact tracing apps (right) over time.}
    \label{fig:interaction_immunity}
\end{figure}

\subsubsection{Influence of Factors on the Willingness to Use COVID-19 Apps Over Time in Germany}

As in the US analysis, for Germany we also differentiate between factors with consistently significant influence and those that were significant in some survey round(s) but not in others.

\paragraph{Consistently influential factors}
Our CLMM model for Germany shows that almost all app scenario factors that had continuous significant influence on participants' willingness to use an app in the US also had such consistent influence over all rounds in Germany, though the effect size (estimates) and significance level (p-values) varied: Both factor levels ``contact tracing'' and ``reduced battery life'' again revealed a continuous positive and negative influence, respectively (estimates 0.27, $ p < 0.01 $ and -0.09, $ p < 0.01 $). 
Compared to the US, \Cref{fig:interaction_immunity} shows that even though contact tracing only had a continuous significant influence across all three survey rounds, we observe noticeable differences between both countries. During survey rounds R1 and R3, the mean willingness to use an app is higher in Germany, which we attribute to the general availability of COVID-19-related apps mainly provided by the federal government in Germany, and thus aligns with the actual usage of COVID-19 apps (see \Cref{tab:demographics}). Conversely, in survey round~2, both countries exhibit comparable mean values for the app purpose ``contact tracing''. As discussed in \Cref{sec:w-to-use-rq1}, we hypothesize that this observation, \ie, lower willingness to use apps in R2 in Germany, correlates with the governmental enforcement of a stringent lockdown caused by a rapid increase in COVID-19 infection rates (see \Cref{fig:incidencelevel}), resulting in an overall reduced perceived utility of contact tracing apps due to limited social contacts.

Any app used to (uniquely) identify an individual also negatively impacted participants' willingness to use it over all survey rounds.
As in the US, the same reduced willingness to use any COVID-19 app emerged for certain app data receivers, namely private companies (estimate -0.19, $ p < 0.01 $) and law enforcement (estimate -0.26, $ p < 0.01 $). Unlike in the US, the public serving as the data receiver consistently had a negative influence over all three survey rounds (estimate -0.20, $ p < 0.01 $). Moreover, apps with the societal implication ``faster update of local COVID-19 regulations'' had a significant impact in all rounds in Germany (estimate 0.12, $ p < 0.01 $). 
German participants' IUIPC scores across both the dimensions Collection (estimate 0.11, $ p < 0.01 $) and Awareness (estimate 0.05, $ p < 0.05 $) also showed a positive influence on the willingness to use any app in all three rounds.

\paragraph{Factors with changing influence over time}
Similar to the US data, the influence of most factors from app scenarios did not change over time. One notable exception is the app purpose ``health certificate'' in R3, in which German participants' willingness to use a health certificate app significantly increased ($ p < 0.01 $; estimate: 0.29), as shown in \Cref{fig:interaction_immunity}. This development coincides with the increased availability of vaccines at the time of R3 in Germany and ongoing discussions in the media about privileges for vaccinated people and the planned introduction of digital vaccine passports.

Beyond app scenarios, the model for Germany in \Cref{tab:clmm_appendix} also shows significant interactions between the survey round and  participants' experience with COVID-19. In R2 and R3 participants who had already tested negative were significantly more likely to install an app than in R1 ($ p < 0.001 $ for both interactions, estimates 0.64 in R2 and 0.61 in R3). At the same time those who had not been tested but suspected to already have been infected were less likely to use an app in R2 ($ p < 0.05 $, estimate -0.27); this interaction was not significant in R3. A likely explanation for the increase in the willingness to use an app for already tested participants is that the Corona-Warn-App was not yet available at the time of the first survey round. By contrast, at the time of R3, it had been upgraded with the ability to register one's COVID-19 tests and quickly receive the results digitally. This could hint at people who were willing (or required) to get tested considering the app a good digital support tool. For participants who assumed they had already been infected and had not yet been tested this functionality did not provide any use cases.

The model also shows a significant interaction for education level over time. Medium education level (high school or associate degree) in interaction with the second survey round negatively impacted participants' willingness to use an app (compared to R1). This effect was not present in R3 and might be due to increased app skepticism in this group during the winter 2020 wave. We also observed a significant and positive interaction between people's satisfaction with the battery life of their phone and their willingness to use an app in R3. This survey round took place in May 2021, with fewer restrictions in place compared to the time of R2, so people might have spent more time outside their homes, rendering the battery life of their phone more important. 

Further, participants who found pandemic measures \emph{too lenient} were more likely to install an app in R1, but less likely to do so in later survey rounds.
People who found the measures against the spread of the coronavirus \emph{about right} were more willing to use an app in R2 than in R1. We did not observe this effect in R3. 
This reflects that more effective methods against the spread of COVID-19, namely vaccines, had become available by the time of survey round~3. 

\textbf{Summary RQ\,2:} We identified factors that impact people's willingness to use COVID-19-related smartphone apps and significantly change over the course of the first 1.5 years of the pandemic, including increased acceptance of health certificate apps in Germany. Other factors consistently remain significant across all three survey rounds in both countries, including the app purpose contact tracing, unique identification of individuals or transmission of their demographic data, and private companies or law enforcement as data recipients. Factors related to the data processing practices of apps less frequently have significant influence on people's willingness to use them compared to individual experience with the pandemic and stance towards certain public and private organizations.

\subsection{Perception of COVID-19 Apps Over Time (RQ\,3)}
\label{sec:qualitative}

Our qualitative analysis of the open-ended responses in our questionnaire provides insights how participants perceived COVID-19 apps over the course of the first 1.5 years of the pandemic.

\subsubsection{Reasons Not to Use COVID-19 Apps}
\label{sec:whynoapp}

\begin{figure*}[t]
	\centering
    \includegraphics[width=\textwidth]{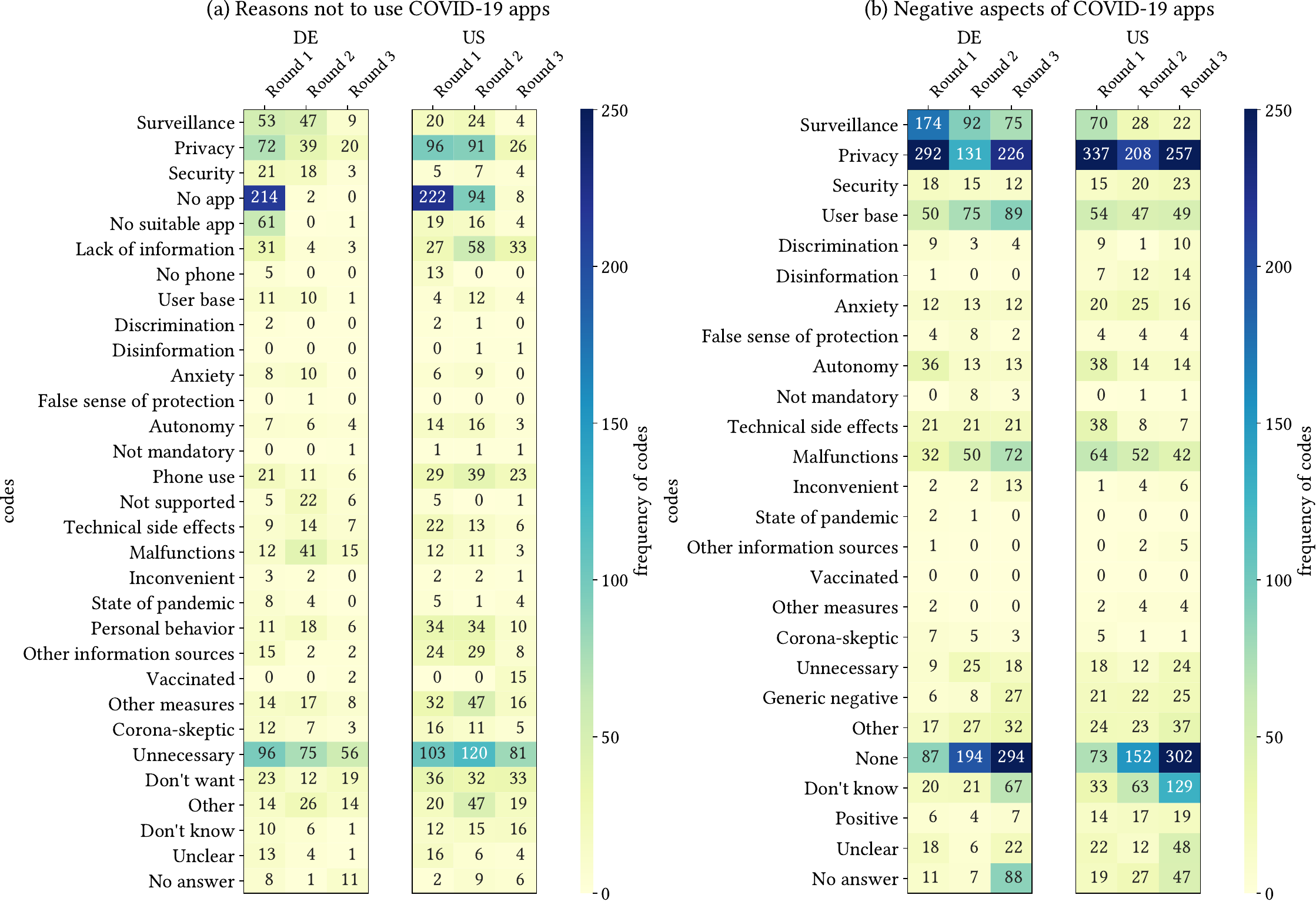}
     \caption{Individual reasons not to use COVID-19 apps ((a); Q21) and perceived negative aspects of COVID-19 apps ((b); Q12) by country and across rounds, absolute numbers.}
	\label{fig:whynoapp-negative-heatmaps}
\end{figure*}

We found participants' reasons to delete previously downloaded apps (Q23) to be a subset of those not to use COVID-19 apps in the first place (Q21), so we applied the same codebook. Further, the answers to these questions were similar to perceived negative aspects of such apps (Q12), leading to codebooks with a large overlap, shown as the y-axis labels in \Cref{fig:whynoapp-negative-heatmaps} (a) and (b). The codebook for reasons to not (or no longer) use an app comprised 31 codes, that for negative aspects 26. Unique to the former were reasons rooted in individual behavior availability, and statements that merely expressed unwillingness to use them without providing any reason. Only the codebook for negative aspects contained codes for generic negative statements and ``none.'' Vaccination was added as a new reason to both codebooks only in Round~3. 
The full codebooks with examples for each code are available in \Cref{app:codebooks}. 
For all open-ended questions, each answer was assigned one or multiple codes. \Cref{fig:whynoapp-negative-heatmaps} shows how often each code was assigned. 
For the codebook for reasons not to use an app or delete a previously installed app, the weighted Krippendorff's alpha value to determine inter-rater agreement (see \Cref{sec:method-qualitative}) was 0.90. Values for individual codes ranged from 0.52 for the ``unclear'' code to 1.

For the reasons not to use a COVID-19 app in the first place (Q21; \Cref{fig:whynoapp-negative-heatmaps} (a)) we received a varying number of open-ended responses over time as the availability of COVID-19 apps in the respective country and consequently the number of app users changed over time: in Germany 685, 310, and 412 answers in Rounds~1--3, and in the US 702, 634, and 786, respectively. 
Correspondingly, we observe a drop in the prevalence of there being ``no app'' or ``no suitable app'': Numbers dropped to almost zero in Germany in Rounds~2 and 3 after the Corona-Warn-App had become available, while in the US we observe a drop by approximately 50\,\% in R2, when participants had also started to name state-specific COVID-19 apps in Q20 (see \Cref{sec:whichapp}). At the same time, with availability of the Corona-Warn-App (see \Cref{sec:germany}) the number of German participants who referred to technical issues as the reason not to adopt a COVID-19 app increased in R2, as hands-on experience with COVID-19 apps had become more widespread.

Across all rounds in both countries, the predominant reason not to use COVID-19 apps were generic statements of an app not being necessary or useful to fight the ongoing pandemic: ``don't see how it will help'' (R1-US-4), ``not useful'' (R2-US-623, R1-DE-791), ``don't need it'' (R3-DE-272, R3-US-65), ``useless app'' (R2-DE-120), though the number of participants with this sentiment decreased between rounds, except for US participants in R2. 
In a similar decline in both countries were other frequently named reasons for non-use: fear of government surveillance (``state spying on citizens'' [R3-US-130], ``Stasi methods'' [R2-DE-95], ``Deep State'' [R1-US-950]), privacy (``too invasive'' [R3-US-20], ``my location is nobody's business'' [R2-DE-280]), and security aspects (``too insecure'' [R2-DE-527], ``can be hacked'' [R1-US-397]). It appears that as time progressed and others had experience with such apps, people's opinion towards them grew less adverse.

\subsubsection{Reasons to Delete COVID-19 Apps}
\label{sec:whydeleted}

In Q23 the relatively small number of participants who indicated to have deleted a previously used COVID-19 app (Round~2: 64 in DE, 17 in US; Round~3: 107 in DE, 27 in US) were asked about the reasons for this. Answers were provided by 64 participants in Round 2 (DE: 53, US: 11) and 133 in Round 3 (DE: 106, US: 27), with higher numbers in Germany again reflecting the availability of national, government-backed COVID-19 apps. There, people most frequently referred to technical issues, particularly malfunctions, mentioned 12 and 8 times in Rounds~2 and 3, respectively (``unreliable'' [R2-DE-695],  ``too many incorrect alerts'' [R2-DE-156], ``no updates for days'' [R2-DE-611], ``it always crashed'' [R3-DE-326], ``was not consistent with regional data'' [R2-DE-495]), and undesirable technical side effects, mentioned 17 and 4 times (``quickly drains the battery'' [R2-DE-78, R2-DE-838, R3-DE-373], ``Bluetooth needs to be turned on at all times'' [R2-DE-21], ``slowed down the phone'' [R2-DE-748]). Second most prevalent (R2-DE: 7, R3-DE: 10; R2-US: 3, R3-US: 8) were generic statements that the app was unnecessary (``obsolete'' [R2-DE-272], ``don't need it anymore'' [R3-US-633], ``it didn't help'' [R3-US-367]). In particular, US participants in R3 (6 out of 27) perceived the app as no longer necessary due to vaccination (``I deleted the app once I'd been fully vaccinated'' [R3-US-23], ``I got my shot'' [R3-US-834]). This hints at misconceptions of the functionality and benefits of COVID-19 apps.

\subsubsection{Perceived Negative Aspects of COVID-19 Apps}
\label{sec:negative}

In Q12 all participants were asked which aspects they generally found negative about smartphone apps designed to help fight the pandemic. The resulting codebook, shown in \Cref{fig:whynoapp-negative-heatmaps} (b) and \Cref{app:codebooks}, contained 26 codes. Its weighted Krippendorff's alpha value was 0.94, with values for individual codes ranging from 1 to 0.49.

As with the reasons not to use an app, over the course of the first 1.5 years of the pandemic we observe a steady decrease in the number of people who worried about government surveillance and their personal autonomy. For privacy, this holds true only between R1 and R3, whereas both countries exhibited their lowest number of participants with privacy worries in R2, when infection rates were about to rise to new all-time highs in both countries and lockdowns were in place throughout Germany.
Note that this effect was not observed in the reasons to not use an app (see \Cref{sec:whynoapp}). 
Here, privacy still was still one of the most frequently named reasons but was overshadowed by the notion that the app was unnecessary, \ie, notions of individual utility.

In Germany, where national government-backed COVID-19 apps had been available since Round~2, we find both an increase in the number of participants who mentioned malfunctions (``false alarms'' [R3-DE-815], ``outdated information'' [R2-DE-3]) as well as aspects revolving around an app's user base (``incorrect information supplied by users'' [R3-DE-408], ``many do not enter their positive test result'' [R2-DE-408]). By contrast, in the US, these numbers slowly decreased across survey rounds. An explanation could be that widespread or prolonged use of an app made people more aware of its drawbacks in everyday use.

Across rounds, we also find an increase in the number of participants who stated that apps designed to help fight the COVID-19 pandemic had no negative aspects. 
This was also the case for perceived positive aspects (see \Cref{fig:positive-idealapp-heatmaps} (a)), so it appears that over the course of the pandemic people became more indifferent in their stance towards COVID-19 apps.

\subsubsection{Perceived Positive Aspects of COVID-19 Apps}
\label{sec:positive}

For positive aspects of COVID-19 apps, we assigned 24 different codes as shown in \Cref{fig:positive-idealapp-heatmaps} (a) and \Cref{app:codebooks}. The weighted Krippendorff's alpha for this codebook was 0.91, with individual positive values ranging from 1 to 0.63. The code ``healthcertificate'' produced a negative alpha value due to it being relatively rare and used by only one coder in the data used to compute inter-rater agreement.

As just mentioned in \Cref{sec:negative}, we found an increasing number of people who perceived COVID-19 apps to have no positive aspects, which adds to the evidence of growing indifference towards such apps. Between R1 and R2 we also observe a noticeable drop in the number of people who provided generic statements that these apps were a useful tool against the pandemic (``it will keep the pandemic down'' [R3-US-230], ``it can slow down the spread'' [R1-DE-263], ``the pandemic is over sooner'' [R2-DE-985]). This could be an effect of the prolonged duration of the pandemic dampening people's expectations of what an app can actually contribute.

Looking at concrete aspects, the use of apps for contact tracing was regarded as positive by a slowly increasing number of participants in Germany, while US numbers steadily declined between rounds. This could be an effect of Germany having issued a national contact tracing app and multiple of its states having used the privately developed app Luca as part of their anti-pandemic strategy (see \Cref{sec:germany}), while such apps have only been rolled out on the state level in the US, resulting in less hands-on experience with them.

\subsubsection{The Ideal COVID-19 App}
\label{sec:idealapp}

\begin{figure}[t]
	\centering
    \includegraphics[width=0.8\textwidth]{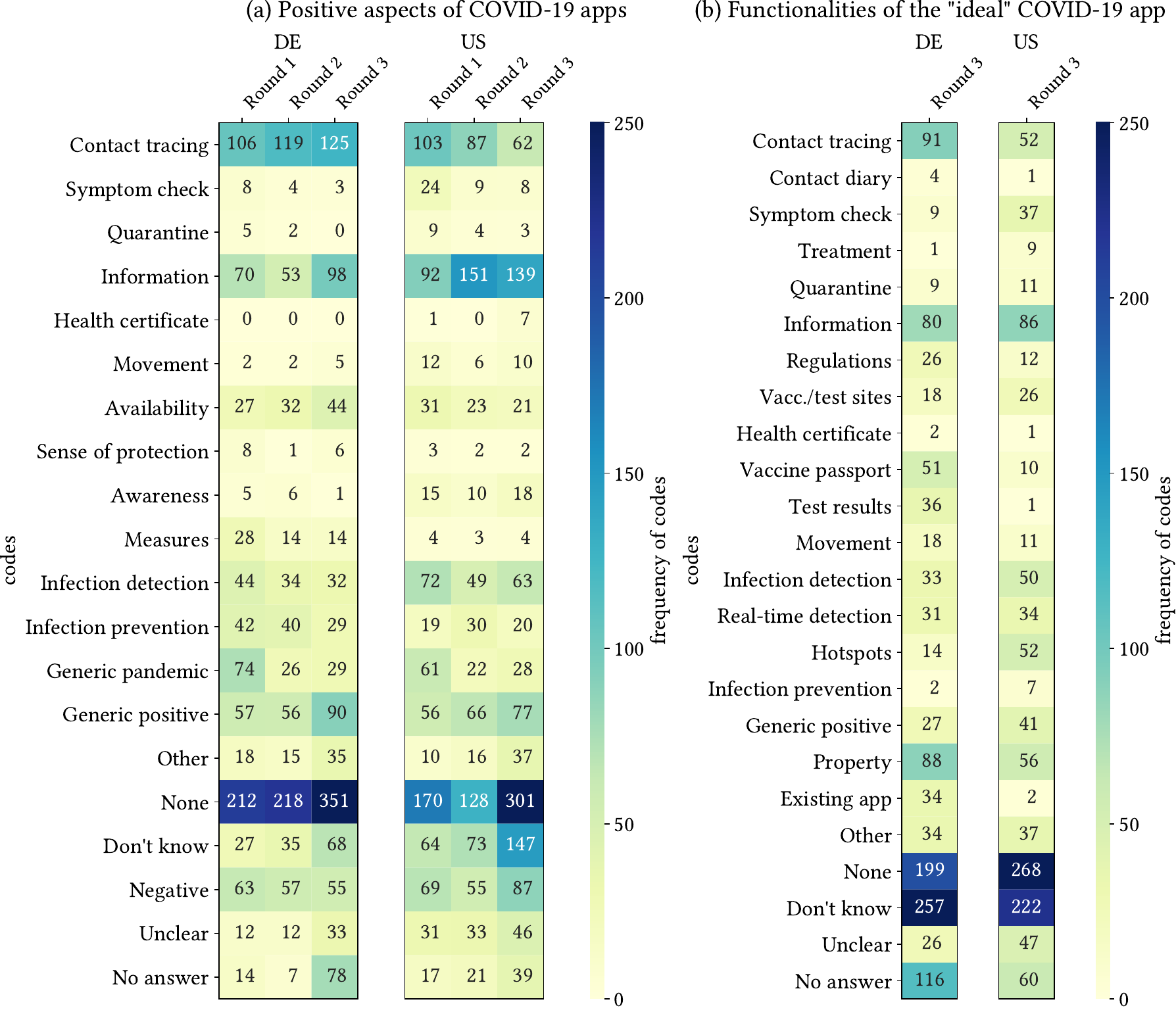}
     \caption{Perceived positive aspects of COVID-19 apps, across all rounds ((a); Q11) and, for Round~3, functionalities participants considered to be part of the ``ideal'' app to help fight the COVID-19 pandemic (Q13); both by country and in absolute numbers.}
	\label{fig:positive-idealapp-heatmaps}
\end{figure}

Since many participants in R2 expressed dissatisfaction with existing COVID-19 apps, we added Q13 in R3 that asked which functionalities they desired in smartphone apps designed to help fight the pandemic. We received 1026 answers from participants in Germany and 1052 from the US. Many answers described functionality already found and perceived to be positive in existing apps, so we used the codebook for perceived positive aspects as a starting point, adding new (sub)codes that occurred frequently\footnote{We did not add these new codes to the codebook for positive aspects (Q11) to preserve comparability across rounds.} and removing those that did not relate to an app's functionality.
The final codebook, shown as the y-axis labels in \Cref{fig:positive-idealapp-heatmaps} (b) and in \Cref{app:codebooks}, comprised 24 codes. The weighted Krippendorff's alpha for this codebook was 0.88, with individual positive values ranging from 0.45 for the ``unclear'' code to 1.

The high number of ``Don't know'' answers (DE: 257, US: 222) indicates that participants found this question difficult to answer. Another large share of participants (DE: 199, US: 268) provided negative sentiments towards COVID-19 apps (``none'' [R3-US-5], ``one to uninstall'' [R3-DE-14], ``I don't want an app'' [R3-DE-274], ``rubbish'' [R3-US-67]), which we coded as the ideal COVID-19 app being ``None.'' As with perceived positive aspects, we found patterns of participants desiring functionality they were already familiar with from existing apps: contact tracing in both countries, and vaccine passports and functionality to manage test results in Germany, where another 34 participants directly referred to the functionality of existing apps.

Some participants also wished for functionality that, with the current state of technology, is impossible to provide by a smartphone app. Most notable (DE: 31, US: 34) was the wish for real-time detection of the virus or infected people in proximity to the app user: ``alert of any confirmed infected people that may be within a certain radius'' (R3-US-611), ``alert me if a sick person is moving towards me'' (R3-DE-128), ``location of positive cases'' (R3-US-1038). A smaller number of participants, particularly in the US, wished for direct treatment (``treat and diagnosing the virus like a doctor'', R3-US-142) or administration of a vaccine via the app (``it would shoot out the vaccine'', R3-US-118; ``it should be able to vaccinate'', R3-DE-485).

Another notable share of participants (DE: 88, US: 56) referred to specific properties of an app rather than functionalities, including ``anonymity'' (R3-DE-302, R3-US-778), ``privacy'' (R3-DE-15, R3-US-1027), ``speed'' (R3-US-623, R3-DE-353), ``works without GPS'' (R3-DE-650), ``easy to understand'' (R3-US-146), ``report the truth, not bias'' (R3-US-913), or ``all people should be required to use it'' (R3-DE-975).
Overall, we find evidence that people have a hard time imagining what is possible with current technology apart from what they are already familiar with.

\textbf{Summary RQ\,3:} We find that, over time, the prolonged availability of COVID-19 apps led people to adopt less extreme opinions of them. While privacy is of continuous importance in the abstract perception of apps, times of severe limitation of personal freedoms yielded fewer reports, and as an argument against concrete app use, privacy is consistently overshadowed by considerations of limited utility. Misconceptions, such as apps no longer being necessary after vaccination, can lead to people stopping their use too early.

%% file: tables/all123_participants.tex
\begin{table*}[ht]
\caption[Participant demographics]{\label{tab:demographics}
Participant demographics, with population statistics as reported by our panel provider.}


\centering
\small

\begin{minipage}{\textwidth}
\renewcommand{\thefootnote}{\thempfootnote}
\renewcommand\footnoterule{}
\makeatletter
\newcommand\footnoteref[1]{\protected@xdef\@thefnmark{\ref{#1}}\@footnotemark}
\makeatother

\newcommand*{\tabmeansd}[1]{\multicolumn{1}{@{}c@{}}{\tablenum[parse-numbers = true, table-format = 2.2(3), separate-uncertainty]{#1}}}

\begin{tabular*}{\textwidth}{
@{}
l
@{\extracolsep{\fill}}
*{3}{S[table-format = 2.2]}
@{\extracolsep{0pt}}H@{\extracolsep{\fill}}
*{3}{S[table-format = 2.2]}
@{\extracolsep{0pt}}Z@{\extracolsep{\fill}}
}
\toprule
& \multicolumn{4}{c}{\textbf{Germany}} & \multicolumn{4}{c}{\textbf{United States}} \\

\cmidrule{2-5}
\cmidrule{6-9}

& \multicolumn{1}{c}{\textbf{Round 1}} & \multicolumn{1}{c}{\textbf{Round 2}} & \multicolumn{1}{c}{\textbf{Round 3}} & \multicolumn{1}{H}{\textbf{Pop.}} & \multicolumn{1}{c}{\textbf{Round 1}} & \multicolumn{1}{c}{\textbf{Round 2}} & \multicolumn{1}{c}{\textbf{Round 3}} & \multicolumn{1}{Z}{\textbf{Pop.}} \\

& \multicolumn{1}{c}{(n = 1003)} & \multicolumn{1}{c}{(n = 1018)} & \multicolumn{1}{c}{(n = 1028)} & & \multicolumn{1}{c}{(n = 1003)} & \multicolumn{1}{c}{(n = 1000)} & \multicolumn{1}{c}{(n = 1072)} & \\

\midrule
\textbf{Gender} & \multicolumn{1}{@{}c}{\textbf{\%}} & \multicolumn{1}{c}{\textbf{\%}} & \multicolumn{1}{c}{\textbf{\%}} & \multicolumn{1}{H}{\textbf{\%}}  & \multicolumn{1}{c}{\textbf{\%}}& \multicolumn{1}{c}{\textbf{\%}} & \multicolumn{1}{c}{\textbf{\%}} & \multicolumn{1}{Z}{\textbf{\%}} \\

Female & 50.65 & 50.29 & 51.75 & 50.19 & 53.04 & 56.60 & 54.01 & 50.75 \\
Male   & 49.35 & 49.71 & 48.25 & 48.81 & 46.96 & 43.40 & 45.99 & 49.25 \\

\midrule
\textbf{Age} & \multicolumn{1}{@{}c}{\textbf{\%}} & \multicolumn{1}{c}{\textbf{\%}} & \multicolumn{1}{c}{\textbf{\%}} & \multicolumn{1}{H}{\textbf{\%}} & \multicolumn{1}{c}{\textbf{\%}} & \multicolumn{1}{c}{\textbf{\%}} & \multicolumn{1}{c}{\textbf{\%}} & \multicolumn{1}{Z}{\textbf{\%}} \\

18--35 & 26.12 & 26.92 & 29.86 & 26.79 & 19.84 & 23.20 & 14.83 & 33.00 \\
36--55 & 39.18 & 39.00 & 40.27 & 37.05 & 45.76 & 43.50 & 42.72 & 39.00 \\
56--80 & 34.70 & 34.09 & 29.86 & 36.16 & 34.40 & 33.30 & 42.44 & 28.00 \\

\midrule
\textbf{Education} & \multicolumn{1}{@{}c}{\textbf{\%}} & \multicolumn{1}{c}{\textbf{\%}} & \multicolumn{1}{c}{\textbf{\%}} & \multicolumn{1}{H}{\textbf{\%}} & \multicolumn{1}{c}{\textbf{\%}} & \multicolumn{1}{c}{\textbf{\%}} & \multicolumn{1}{c}{\textbf{\%}} & \multicolumn{1}{Z}{\textbf{\%}} \\
        
Primary School
&  1.30 &  1.87 &  2.04 &  &  1.40 &  0.40 &  1.21 &  \\
Secondary School
&  3.19 &  6.29 &  5.84 &  &  5.98 &  5.00 &  3.64 &  \\
HS/Tertiary/Tech. College
& 60.02 & 71.61 & 70.53 &  & 40.58 & 38.80 & 39.09 &  \\
University/Higher Ed.\footnote{\label{footnote:degree-germany}We do not differentiate between undergraduate and (post)graduate degrees in Germany as this was only introduced in the 2000s.}
& \multirow{2}{=}{23.83} & \multirow{2}{=}{20.14} & \multirow{2}{=}{21.40} &  & 42.37 & 45.00 & 43.00 &  \\
Postgraduate Education\footnoteref{footnote:degree-germany}
&       &       &       &  & 9.67 & 10.80 & 13.06 &  \\
Other
& 11.67 &  0.10 &  0.19 &  &  \multicolumn{1}{c}{-} &   \multicolumn{1}{c}{-} &   \multicolumn{1}{c}{-} &  \\
        
\midrule
\textbf{COVID-19 Apps} & \multicolumn{1}{@{}c}{\textbf{\%}} & \multicolumn{1}{c}{\textbf{\%}} & \multicolumn{1}{c}{\textbf{\%}} & \multicolumn{1}{H}{\textbf{\%}} & \multicolumn{1}{c}{\textbf{\%}} & \multicolumn{1}{c}{\textbf{\%}} & \multicolumn{1}{c}{\textbf{\%}} & \multicolumn{1}{Z}{\textbf{\%}} \\
Use of an app & 3.99 & 38.90 & 42.70 & \multicolumn{1}{H}{-} & 6.18 & 11.30 & 8.58 & \multicolumn{1}{Z}{-} \\

\midrule

\textbf{IUIPC} & \multicolumn{1}{c}{\textbf{mean (sd)}} & \multicolumn{1}{c}{\textbf{mean (sd)}} & \multicolumn{1}{c}{\textbf{mean (sd)}} & & \multicolumn{1}{c}{\textbf{mean (sd)}} & \multicolumn{1}{c}{\textbf{mean (sd)}} & \multicolumn{1}{c}{\textbf{mean (sd)}} & \\


Control    & \tabmeansd{6.16(120)} & \tabmeansd{5.82(139)} & \tabmeansd{5.75(140)} & & \tabmeansd{5.51(124)} & \tabmeansd{5.58(116)} & \tabmeansd{5.49(127)} & \\
Awareness  & \tabmeansd{5.43(132)} & \tabmeansd{5.32(142)} & \tabmeansd{5.21(143)} & & \tabmeansd{6.04(118)} & \tabmeansd{6.06(111)} & \tabmeansd{5.98(123)} & \\
Collection & \tabmeansd{5.67(117)} & \tabmeansd{5.47(129)} & \tabmeansd{5.36(129)} & & \tabmeansd{5.64(129)} & \tabmeansd{5.61(126)} & \tabmeansd{5.64(134)} & \\

\bottomrule
\end{tabular*}
\end{minipage}
\end{table*}

%% file: tables/deus123_clmm_shortened_v2.tex
\definecolor{col-hline}{RGB}{184,84,80}
\definecolor{col-us}{RGB}{248,206,204}
\definecolor{col-de}{RGB}{248,206,204}

\small

\begin{table}[h]
\caption{\label{tab:clmm_shortened}
Estimates from the cumulative link mixed models (CLMMs) for participants’ willingness to use COVID-19 apps for survey round, app scenario factors (\eg, app purpose), and non-app scenario factors (IUIPC). The statistics for factors that are consistently significant in all survey rounds are \factor{red}, \ie, factors that keep having a significant (positive or negative) impact on the willingness to use apps over time.
The \factorrounds{blue} suffixes R2 and R3 indicate a significant change compared to round R1, \ie, a change in interaction with the first survey round. Significance levels are indicated with stars (*$p<.05$, **$p<.01$).
}

\begin{tabular}{@{}l@{}*{2}{c}@{}}

\toprule
\textbf{Factors + Levels} & \textbf{United States}&\textbf{Germany} \\

\midrule

\multicolumn{3}{@{}l}{\textbf{Survey Round (baseline: Round 1)}}\\
Round 2&-1.41** & 0.51* \\

Round 3&-0.86* & 0.95** \\

\midrule
\emph{App scenario factors}&&\\
\midrule
\multicolumn{3}{@{}l}{\textbf{Purpose (baseline: information)}}\\

Symptom check&0.14&0.04\\

\factor{Contact tracing}&\factor{0.21*}&\factor{0.27**}\\

Quarantine enforcement\phantom{-----------------}&0.04&0.08\\

Health certificate&0.07&0.07\\

\factorrounds{Health certificate:R3}&-&\factorrounds{0.29**}\\

\midrule \multicolumn{3}{@{}l}{\textbf{Technical implications (baseline: none)}}\\
\factor{Reduced battery life}&\factor{-0.12*}&\factor{-0.09**}\\

\midrule \multicolumn{3}{@{}l}{\textbf{Societal implications (baseline: none)}}\\
Faster update of regional rules&0.04&\factor{0.12**}\\

\midrule \multicolumn{3}{@{}l}{\textbf{Payload data (baseline: none)}}\\
Location&-0.06&0.04\\
\factorrounds{Location:R2}&\factorrounds{0.14*}&-\\

\factorrounds{Location:R3}&\factorrounds{0.15*}&-\\

\midrule \multicolumn{3}{@{}l}{\textbf{Identification data (baseline: none)}}\\
\factor{Demographic data}&\factor{-0.18*}&\factor{-0.20**}\\
\factor{Unique identification of individual}&\factor{-0.25**}&\factor{-0.20**}\\

\midrule \multicolumn{3}{@{}l}{\textbf{Data receiver (baseline: unspecified)}}\\
\factor{Private companies}&\factor{-0.18*}&\factor{-0.19**}\\
\factor{Law enforcement}&\factor{-0.25**}&\factor{-0.26**}\\

\midrule
\emph{Non-scenario factors}&&\\
\midrule

\multicolumn{3}{@{}l}{\textbf{[Q30] IUIPC}}\\

\factor{Collection}&\factor{-0.13**}&\factor{0.11**}\\

\factorrounds{Collection:R2}&\factorrounds{0.09*}&-\\

\factorrounds{Collection:R3}&\factorrounds{0.10*}&-\\

Awareness&0.02&\factor{0.05*}\\

\factorrounds{Awareness:R2}&\factorrounds{-0.16**}&-\\

\factorrounds{Awareness:R3}&\factorrounds{-0.12*}&-\\
\bottomrule

\end{tabular}
\end{table}
\normalsize

%% file: sections/5_discussion.tex
\section{Discussion}
\label{sec:discussion}

Our results indicate that the influence of multiple investigated factors on the adoption of COVID-19 apps changed during the first 1.5 years of the COVID-19 pandemic, whereas other factors consistently affected people’s willingness to use such apps over all three survey rounds. 
We now discuss the implications of our findings for the design of future apps that serve similar purposes beyond the COVID-19 pandemic. 
In particular, as the factor levels used in the app scenarios were originally derived from real-world COVID-19 apps issued or backed by national governments~\cite{utz_covidapps_2021}, our findings can inform the design of future apps issued by public actors that collect and process users' personal information for the benefit of society as a whole and would benefit from widespread voluntary adoption. 
Possible use cases could be apps to tackle future health crises on a more general scale, for the information and management of natural disasters like tornadoes and floods, or that simply collect citizen-provided data to inform projects to improve local or national infrastructure.

\subsection{The Continuous Importance of Data Privacy}

Our analyses revealed multiple factors of continuous importance in people's willingness to adopt COVID-19 apps and how they perceive such apps, most of which relate to data privacy.

In our CLMM models we found less privacy-preserving implementations to consistently reduce people's willingness to use an app. For example, we found apps that transmit data that allows for unique identification of the individual or even only collect demographic information to continuously have negative impact on both US and German participants' willingness to use a COVID-19 app. 
Note that the role of user anonymity is an aspect that depends on cultural norms, as our earlier work~\cite{utz_covidapps_2021} has found the exact opposite effect for China, where identifiability of the individual had a significant positive influence on people's willingness to use a COVID-19 app.

In the same vein, sharing data with third parties who are not directly concerned with pandemic control (\ie, private companies or law enforcement) also significantly reduced people's willingness to use COVID-19 apps.
Participants were also more willing to use apps for less invasive purposes, \eg, apps for contact tracing and information as opposed to apps enforcing quarantine or for symptom checks that prompted for sensitive health information.

Combining these observations we can deduce that people expect the data flows of an app to be appropriate to the context in which they occur, thus confirming Nissenbaum's theory of privacy as contextual integrity~\cite{nissenbaum_privacy_2004}. 

These findings are supported by the open-ended survey responses, with privacy being the most frequently mentioned negative aspect of COVID-19 apps in both countries and even a frequent reason not to use COVID-19 apps at all.
The still comparably high adoption of actual apps in Germany can also be explained by the theory of CI:
While in a general context, privacy has high value for a participant, in a specific situation, such as a global pandemic in which some actors tie fundamental rights to app use, its importance can be re-evaluated.

This also explains our finding that in both countries privacy concerns were more frequently voiced as negative aspects of COVID-19 apps in the first and third survey rounds than in the second. 
The pandemic state around R2 coincided with the highest incidence rate across all three survey rounds and severe contact and travel restrictions in an attempt to slow down rising infection rates, particularly in Germany.
This again implies that people may still consider privacy important but are more willing to trade it in for a greater good under extraordinary circumstances, temporarily adjusting what they consider ``appropriate'' data flows.

Taking all of this into account, we recommend that future state-issued apps for widespread adoption follow a ``privacy by design'' approach and only collect the amount of data that is necessary to implement the desired functionality. As the discussion of different architectures for digital contact tracing has shown, such requirements can lead to the development of innovative new technology that unites public actors' wish to collect information and citizens' legitimate privacy concerns.

\subsection{Changing Influences: External Factors and Experience with COVID-19 and Apps} 

We also identified factors whose influence on people's willingness to adopt a COVID-19 app changed over time. 
Looking at commonalities between these factors, we find that most relate to external events that are subject to change over a pandemic and have nothing to do with the inherent properties of an app: a person's individual experience with the disease, such as test results or suspected past infections; how they rate measures taken in their community for pandemic containment and control, and their opinion of actors in the pandemic, including health authorities and federal and local governments. 
The latter directly relates to the importance of trust in times of public crisis (see \Cref{sec:discussion-trust}). 

For app designers, such externalities are hard to impossible to predict and consider in the app design process, and updates that try to accommodate such changes may come too late. 

It appears more attainable to consider various events that may occur in the type of crisis situation that fits the use case of an app and try to account for changes in user stance and behavior that these events may objectively cause. One such example would be the effect of lockdowns or stay-at-home orders: If people stay home more, they have fewer use cases for apps for digital contact tracing that record their recent encounters with other people or health certificate apps that provide access to certain locations or services. Apps could display information that point out the importance of their continued use (see also \Cref{sec:discussion-benefits}) to encourage people to keep using them despite temporarily changed circumstances.

Beyond such externalities, we found that the actual availability of apps and experience with them, be it first-hand or reported, naturally changed over time and had major influence on people's stance towards and the reported and actual willingness to adopt COVID-19 apps.

The actual use of COVID-19-related smartphone apps in Germany continuously increased over the first 1.5 years of the pandemic (up to 43\,\% in R3), while in the US the adoption rate remained comparatively low (highest at 11\,\% in R2). 
Accompanying these developments in actual app use, our analysis in \Cref{sec:clmmmodel} found for both countries similar changes in the influence of the three survey rounds on people's willingness to use (hypothetical) apps (see~\Cref{tab:clmm_appendix}).
While in Germany the willingness to use apps was significantly higher in~R2 and~R3 compared to~R1, we observe the exact opposite effect in the US---later survey rounds had significantly negative effects on people's willingness to use COVID-19 apps.
We assume that these diverging developments can be partially attributed to the availability of a federal COVID-19 app in the respective country.
While in Germany the government-backed Corona-Warn-App was widely used, there was no comparable contact tracing solution available on the federal level in the US, which may be a reason for comparably low adoption in practice. 
A lack of hands-on experience with a given technology and its benefits may negatively impact people's stance towards it and, subsequently, their willingness to use it. 
Again, timely open public discussion of apps to be issued by public entities for wide adoption in crisis situations can help familiarize prospective users with their functionality and benefits to foster understanding and the willingness to use them.

Our qualitative analysis found misconceptions among people's negative stance towards COVID-19 apps and suggestions for future functionality, likely caused by incorrect assumptions about the purpose of an app or what is possible with the respective technology.  Missing feedback from the app, such as contact tracing apps only actively changing something in their user interface if a risk encounter was registered, might cause misconceptions or the perception that the app does not work properly.
Kotthaus et al.~\cite{kotthaus_mobilewarningapps_2016} found similar effects for the perception of crisis apps in 2016.

We recommend for the development of future government-issued apps whose efficiency depends on widespread voluntary adoption to more accurately describe the intended purpose and functionality of the app and highlight personal and societal benefits to motivate individuals to use it, but also to clearly communicate its boundaries.
Explaining what an app can and cannot do could avoid raising false expectations that do not come true in the end. 
As we found people's perception of COVID-19 apps to change over time, particularly due to the influence of external effects such as trust in institutions and public policy, it is important that this communication considers such changes and the aforementioned information is not just provided once to motivate app installation but continuously upheld and adjusted throughout the lifetime of an app to clearly communicate the benefits and importance of its continued use.

\subsection{The Role of Trust in Actors and Institutions in Crisis Management}
\label{sec:discussion-trust}

In both the US and Germany, we identified a favorable attitude towards state authorities to have a significant positive impact on participants' willingness to use COVID-19 apps across all three survey rounds. 
As already discussed in more detail in \Cref{sec:rw-crisis-apps}, Dressel~\cite{dressel_riskcultures_2015} has shown in her work conducted before the COVID-19 pandemic that people in high-trust states (\emph{state-oriented risk cultures}) are willing to contribute individually in a crisis, are interested in being integrated in crisis management, and that trust in authorities is an important factor for the use of mobile crisis apps.
For the particular case of the Corona-Warn-App for digital contact tracing released by the German government and health authorities, the research findings of Reuter et al.~\cite{reuter_katwarn_2017} were confirmed in a real crisis: Citizens have a positive attitude toward mobile crisis apps, implying that government-issued apps are trusted in times of crisis. 
On the contrary, data being sent to private companies or law enforcement had a significant negative impact on reported app adoption. Thus, public actors issuing apps designed for widespread adoption in public crises should carefully consider who they partner with for the development of such apps, what third-party components are being used, and who has access to the collected data. The purpose and data processing practices of the app must be clearly outlined beforehand to foster people's trust that the app only does what it claims to do. 
Ideally, the source code is made publicly available to allow for review by security and privacy experts.

Once established, institutions need to continuously work on keeping up users' trust in an app, taking care to not succumb to ``function creep''~\cite{bridewell_functioncreep_2023} and exceed the capabilities and data processing practices they originally promised an app to have. Though we found the influence of trust in institutions to have varying influence on people's willingness to use COVID-19 apps, real-world events such as surrendering contact tracing data from the app Luca to the police~\cite{pannett_germanpolice_2022} have shown that trust in an app can quickly be damaged.
Similarly, ongoing discussions about the future of existing COVID-19 apps and what will happen to the collected data are facing the argument that these apps were initially introduced with the intention of being a tool for a very specific purpose and a limited period of time in a global crisis, which should not be extended or overstepped retroactively, thus again highlighting the importance of trust in an app and the entities behind it.

\subsection{Societal vs. Individual Benefits}
\label{sec:discussion-benefits}

Real-world adoption of COVID-19 apps and related research have shown that people generally are willing to install and use smartphone apps for societal benefit---though not without limits.

First, people's opinions appear to be shaped by what they are already familiar with through media discussions or actual use in their country. 
In our CLMM models only the app purpose contact tracing had a significant positive influence on participants' willingness to use apps over all three survey rounds in both countries, confirming the results of Dressel et al.~\cite{dressel_riskcultures_2015} that people are motivated to provide an individual contribution to help overcome a public crisis.
The same applies for the significant change in participants’ willingness to use health certificate apps in Germany in Round~3, a finding confirmed for actual use of such apps by other work from our research group~\cite{herbert_adoptioncovidcert_2022} who found in a representative study in Germany in December 2021 that 70\,\% of participants actually used an app of this type (digital COVID certificate). 

It has to be noted that the high reported user bases for these apps, particularly health certificates, could partly be due to the individual benefits they granted to users: for digital contact tracing apps, the ability to estimate one's own risk of recent exposure to COVID-19, and for health certificates, the ability to obtain access to locations or services only available to people with a certain health status, such as vaccination or a recent negative test result. 
We found evidence in our data that some people were not or no longer willing to use COVID-19 apps if they did not consider them to provide sufficient individual benefit: 
Across all three survey rounds, the most prominent reason participants named for not using a COVID-19 app was such apps being unnecessary in a generic way, and multiple US participants in R3 reported to have deleted a COVID-19 app because they no longer considered it necessary after having received the vaccine. 
For the development of future apps designed for widespread adoption in emergency or crisis situations this means that designers should keep communicating to users how they could as individuals benefit from (continued) use of an app that was mainly designed to benefit society as a whole, rather than provide individual short-term advantage.

Despite the above observation, there were real-life COVID-19 apps that were downloaded and used even though they did not provide an immediate personal benefit to users. 
This is supported by our CLMM models, in both of which the promise of individual benefits in the ``societal implications'' factor did not have significant influence on participants' decision to use a COVID-19 app in any of the survey rounds. 
One example for a COVID-19 app with no immediate personal benefit was the German Corona-Datenspende (``Corona Data Donation'')~\cite{rki_coronadatenspende_2020}, which allowed people to contribute their health and activity data for scientific research.
As it could only used by people who owned a fitness tracker, its user base was nowhere near as large as those of national contact tracing apps, so there are limited insights into users' decisions for (prolonged) use of the app. 
Future work could investigate under what circumstances people would be willing to adopt apps for purely altruistic purposes.

In the EU and in Germany in particular, public debates sparked about the privacy, security, and societal implications of government-backed health applications during their development phase and in advance of their release. This ultimately led to the more privacy-friendly approach of a decentralized and Bluetooth proximity-based app, \eg, the Corona-Warn-App for digital contact tracing in Germany, which has a comparatively high acceptance.

%% file: sections/7_conclusion.tex
\section{Conclusion}
\label{sec:conclusion}

The COVID-19 pandemic has demonstrated how smartphone apps can make a meaningful contribution in a public (health) crisis and help slow down the spread of certain types of infectious diseases, but at times only when they are widely adopted.
We conducted online surveys in Germany and the United States at three distinct points in time during the first 1.5 years of the COVID-19 pandemic to learn how people's attitudes towards smartphone apps designed to aid the fight against COVID-19 developed over the course of the ongoing pandemic.
Fears of surveillance and perceived risks for privacy, security, and personal autonomy became less pronounced over time while people's adoption of---but also indifference to---such apps increased.
We found external factors such as pandemic state and containment measures to influence the willingness to use COVID-19 apps. 
Still, we identified future opportunities to foster use of these and similar apps:
respecting users' expectations of privacy and the appropriateness of data flows, along with focusing on perceived utility and user understanding of the functionality of an app and the importance of its continued use through transparency and explanations.

%% file: tables/deus123_clmm9.tex
\section{CLMM Models}
\label{sec:clmm_appendix}

\definecolor{col-hline}{RGB}{184,84,80}
\definecolor{col-us}{RGB}{248,206,204}
\definecolor{col-de}{RGB}{248,206,204}

\begin{center}
\small

\begin{longtable}{@{}l@{}*{8}{c}@{}}
\caption{\label{tab:clmm_appendix}
Cumulative link mixed models (CLMMs) for participants’ willingness to use COVID-19 apps (Q14). Highlighted in \factor{red} are the statistics for factors
that are consistently significant in all survey rounds, \ie, factors that continue to have a positive or negative impact on the willingness to use apps over time. The \factorrounds{blue} factors with suffixes R2 or R3 indicate a significant change compared to round R1 for only the respective survey round (\ie, their influence changes in interaction with time).
The table first lists \emph{app scenario factors} that originate in the ten app scenarios shown to participants, followed by \emph{non-scenario factors} from the responses to the survey questions outside the app scenarios plus demographic information.
The most relevant coefficients listed below are estimate (Est.) and \(Pr(>|z|)\). The estimate describes the effect size of this factor within the model. Negative estimates mean that the factor has a negative effect on the reference value (here: the willingness to use an app), while positive values indicate supporting factors. \(Pr(>|z|)\) is equivalent to the p-value. The z-value is the effect size divided by the standard error. Z-values larger than 2 are often interpreted as significant indicators that the estimate is in fact not zero.}
\\

\toprule
\textbf{Factors + Levels} & \multicolumn{4}{c}{\textbf{United States}} & \multicolumn{4}{c}{\textbf{Germany}} \\

\cmidrule{2-5}
\cmidrule{6-9}
& Est. & \footnotesize Std. Err. & z & Pr($>\mid$z$\mid$) & Est. & \footnotesize Std. Err. & z & Pr($>\mid$z$\mid$)\\
\midrule
\endfirsthead

\multicolumn{9}{@{}l}{{\bfseries \tablename\ \thetable{} -- continued from previous page}} \\

\toprule
\textbf{Factors + Levels} & \multicolumn{4}{c}{\textbf{United States}} & \multicolumn{4}{c}{\textbf{Germany}} \\

\cmidrule{2-5}
\cmidrule{6-9}
& Est. & \footnotesize Std. Err. & z &  Pr($>\mid$z$\mid$) & Est. & \footnotesize Std. Err. & z & Pr($>\mid$z$\mid$)\\
\midrule
\endhead

\midrule
\multicolumn{9}{r}{{Continued on next page}} \\
\endfoot

\bottomrule
\endlastfoot

\multicolumn{9}{@{}l}{\textbf{Survey Round (baseline: Round 1)}}\\

Round 2&-1.41&0.38&-3.71&$<$ 0.01&0.51&0.26&-2.00&0.05\\

Round 3&-0.86&0.37&-2.31&0.02&0.95&0.26&-3.67&$<$ 0.01\\

\midrule
\emph{App scenario factors} &&&&&&&&\\
\midrule
\multicolumn{9}{@{}l}{\textbf{Purpose (baseline: information)}}\\

Symptom check&0.14&0.09&1.46&0.14&0.04&0.05&0.78&0.44\\

\factor{Contact tracing}&\factor{0.21}&\factor{0.10}&\factor{2.22}&\factor{0.03}&\factor{0.27}&\factor{0.06}&\factor{4.87}&\factor{$<$ 0.01}\\

Quarantine enforcement&0.04&0.10&0.43&0.67&0.08&0.06&-1.34&0.18\\

Health certificate&0.07&0.12&0.60&0.55&0.07&0.07&-1.08&0.28\\

\factorrounds{Health certificate:R3}&-&-&-&-&\factorrounds{0.29}&\factorrounds{0.09}&\factorrounds{3.128}&\factorrounds{$<$ 0.01}\\

\midrule \multicolumn{9}{@{}l}{\textbf{Technical implications (baseline: none)}}\\

\factor{Reduced battery life}&\factor{-0.12}&\factor{0.06}&\factor{-2.18}&\factor{0.03}&\factor{-0.09}&\factor{0.03}&\factor{-2.85}&\factor{$<$ 0.01}\\

Malfunction contact tracing&-0.06&0.11&-0.53&0.60&0.12&0.06&-1.94&0.05\\

Malfunction information&-0.11&0.11&-1.02&0.31&0.02&0.06&-0.30&0.77\\

Malfunction quarantine enforcement&-0.07&0.11&-0.59&0.56&0.08&0.06&-1.25&0.21\\

Malfunction symptom check&-0.09&0.11&-0.81&0.41&0.02&0.06&-0.32&0.75\\

Malfunction health certificate&-0.17&0.12&-1.46&0.15&0.11&0.06&-1.78&0.07\\

\midrule \multicolumn{9}{@{}l}{\textbf{Societal implications (baseline: none)}}\\

Personal advantages&-0.02&0.09&-0.19&0.85&0.00&0.05&0.07&0.94\\

Faster update of regional rules&0.04&0.07&0.53&0.60&\factor{0.12}&\factor{0.04}&\factor{3.10}&\factor{$<$ 0.01}\\

Future use cases&-0.09&0.07&-1.34&0.18&0.00&0.04&-0.11&0.91\\

\midrule \multicolumn{9}{@{}l}{\textbf{Payload data (baseline: none)}}\\

Location&-0.06&0.05&-1.14&0.25&0.04&0.03&1.24&0.21\\

\factorrounds{Location:R2}&\factorrounds{0.14}&\factorrounds{0.06}&\factorrounds{2.08}&\factorrounds{0.04}&-&-&-&-\\

\factorrounds{Location:R3}&\factorrounds{0.15}&\factorrounds{0.06}&\factorrounds{2.33}&\factorrounds{0.02}&-&-&-&-\\

Health Information&-0.10&0.06&-1.80&0.07&0.05&0.03&-1.69&0.09\\

Infection Status&-0.05&0.06&-0.84&0.40&0.03&0.02&-1.43&0.15\\

Encounters&0.02&0.06&0.31&0.76&0.02&0.04&0.45&0.65\\

Unspecified&-0.03&0.23&-0.13&0.90&0.05&0.13&-0.41&0.68\\

\pagebreak
\multicolumn{9}{@{}l}{\textbf{Identification data (baseline: none)}}\\

\factor{Demographic data}&\factor{-0.18}&\factor{0.07}&\factor{-2.41}&\factor{0.02}&\factor{-0.20}&\factor{0.04}&\factor{-4.66}&\factor{$<$ 0.01}\\

\factor{Unique identification of individual}&\factor{-0.25}&\factor{0.07}&\factor{-3.27}&\factor{$<$ 0.01}&\factor{-0.20}&\factor{0.04}&\factor{-4.77}&\factor{$<$ 0.01}\\

\midrule 
\multicolumn{9}{@{}l}{\textbf{Data receiver (baseline: unspecified)}}\\

Health authorities&-0.08&0.07&-1.02&0.31&0.00&0.04&-0.02&0.99\\

Research institutes&0.03&0.07&0.44&0.66&0.03&0.04&-0.64&0.52\\

\factor{Private companies}&\factor{-0.18}&\factor{0.07}&\factor{-2.40}&\factor{0.02}&\factor{-0.19}&\factor{0.04}&\factor{-4.47}&\factor{$<$ 0.01}\\

\factor{Law enforcement}&\factor{-0.25}&\factor{0.08}&\factor{-3.27}&\factor{$<$ 0.01}&\factor{-0.26}&\factor{0.04}&\factor{-5.93}&\factor{$<$ 0.01}\\

Public&-0.20&0.11&-1.80&0.07&\factor{-0.20}&\factor{0.06}&\factor{-3.17}&\factor{$<$ 0.01}\\

\midrule \multicolumn{9}{@{}l}{\textbf{Retention period (baseline: unspecified)}}\\

One month&-0.06&0.06&-0.94&0.35&0.01&0.03&0.29&0.78\\

End of COVID-19 pandemic&0.01&0.06&0.14&0.89&0.02&0.03&0.57&0.57\\

\midrule
\multicolumn{9}{@{}l}{\textbf{Data transmission (baseline: automatically)}}\\

Manual&0.02&0.04&0.39&0.70&0.03&0.03&1.03&0.30\\

\midrule 
\emph{Non-scenario factors} &&&&&&&&\\

\midrule\multicolumn{9}{@{}l}{\textbf{[Q2]: Phone OS (baseline: Android)}}\\

\factor{iOS}&\factor{0.16}&\factor{0.03}&\factor{5.90}&\factor{$<$ 0.01}&\factor{0.15}&\factor{0.03}&\factor{5.90}&\factor{$<$ 0.01}\\

\midrule \multicolumn{9}{@{}l}{\textbf{[Q3.1]: Satisfied with battery life (baseline: neutral)}}\\

satisfied&-0.04&0.12&-0.33&0.74&0.08&0.06&-1.43&0.15\\

\factorrounds{satisfied:R3}&-&-&-&-&\factorrounds{0.30}&\factorrounds{0.08}&\factorrounds{3.56}&\factorrounds{$<$ 0.01}\\

dissatisfied&-0.27&0.15&-1.85&0.06&0.07&0.08&-0.92&0.36\\

\midrule
\multicolumn{9}{@{}l}{\textbf{[Q3.2]: Satisfied with location accuracy (baseline: neutral)}}\\

satisfied&0.16&0.12&1.38&0.17&\factor{0.29}&\factor{0.05}&\factor{5.59}&\factor{$<$ 0.01}\\

dissatisfied&0.18&0.23&0.78&0.44&\factor{0.68}&\factor{0.18}&\factor{3.76}&\factor{$<$ 0.01}\\

\midrule
\multicolumn{9}{@{}l}{\textbf{[Q3.3]: Satisfied with camera (baseline: neutral)}}\\

satisfied&0.04&0.05&0.79&0.43&0.02&0.04&-0.55&0.58\\

dissatisfied&-0.04&0.08&-0.56&0.57&\factor{0.12}&\factor{0.06}&\factor{-2.02}&\factor{0.04}\\

\midrule \multicolumn{9}{@{}l}{\textbf{[Q3.4]: Satisfied with phone speed (baseline: neutral)}}\\

satisfied&\factor{-0.34}&\factor{0.12}&\factor{-2.73}&\factor{$<$ 0.01}&0.02&0.06&0.31&0.75\\

\factor{dissatisfied}&\factor{-0.50}&\factor{0.21}&\factor{-2.42}&\factor{0.02}&\factor{0.64}&\factor{0.12}&\factor{5.55}&\factor{$<$ 0.01}\\

\midrule
\multicolumn{9}{@{}l}{\textbf{[Q4]: Infected participant (baseline: not tested, no infection suspected)}}\\

Tested negative only&0.09&0.10&0.88&0.38&\factor{0.62}&\factor{0.09}&\factor{-6.96}&\factor{$<$ 0.01}\\

\factorrounds{Tested negative only:R2}&-&-&-&-&\factorrounds{0.64}&\factorrounds{0.10}&\factorrounds{6.38}&\factorrounds{$<$ 0.01}\\

\factorrounds{Tested negative only:R3}&-&-&-&-&\factorrounds{0.61}&\factorrounds{0.10}&\factorrounds{6.36}&\factorrounds{$<$ 0.01}\\

Not tested, infection suspected&0.05&0.13&0.40&0.69&\factor{0.37}&\factor{0.07}&\factor{4.93}&\factor{$<$ 0.01}\\

\factorrounds{Not tested, infection suspected:R2}&-&-&-&-&\factorrounds{-0.27}&\factorrounds{0.11}&\factorrounds{-2.34}&\factorrounds{0.02}\\

\factorrounds{Not tested, infection suspected:R3}&\factorrounds{0.34}&\factorrounds{0.16}&\factorrounds{2.18}&\factorrounds{0.03}&-&-&-&-\\

Tested positive&\factor{0.63}&\factor{0.22}&\factor{2.90}&\factor{$<$ 0.01}&0.07&0.17&-0.38&0.71\\

\factorrounds{Tested positive:R3}&\factorrounds{-0.77}&\factorrounds{0.23}&\factorrounds{-3.33}&\factorrounds{$<$ 0.01}&-&-&-&-\\

Prefer not to answer&\factor{-0.81}&\factor{0.31}&\factor{-2.59}&\factor{$<$ 0.01}&0.14&0.32&0.43&0.67\\

\pagebreak 
\multicolumn{9}{@{}l}{\textbf{[Q5]: Infection in social circles (baseline: no)}}\\

\factor{Yes}&\factor{-0.13}&\factor{0.03}&\factor{-4.09}&\factor{$<$ 0.01}&\factor{0.07}&\factor{0.03}&\factor{2.53}&\factor{0.01}\\

\midrule \multicolumn{9}{@{}l}{\textbf{[Q6]: Quarantine experience (baseline: no)}}\\

Yes&-0.05&0.03&-1.68&0.09&0.06&0.03&1.80&0.07\\

\midrule 
\multicolumn{9}{@{}l}{\textbf{[Q7]: Infection concerns (baseline: somewhat concerned)}}\\

Slightly or not concerned&-0.01&0.04&-0.31&0.76&\factor{0.21}&\factor{0.03}&\factor{-8.03}&\factor{$<$ 0.01}\\

\factor{Moderately or extremely concerned}&\factor{0.40}&\factor{0.04}&\factor{10.68}&\factor{$<$ 0.01}&\factor{0.13}&\factor{0.03}&\factor{5.04}&\factor{$<$ 0.01}\\

\midrule
\multicolumn{9}{@{}l}{\textbf{[Q8]: Person at risk in household (baseline: no)}}\\

Yes&-0.02&0.03&-0.70&0.48&0.04&0.02&-1.65&0.10\\

\midrule \multicolumn{9}{@{}l}{\textbf{[Q10.1] Knows app for COVID-19 info}}\\

\factor{No}&\factor{-0.15}&\factor{0.04}&\factor{-3.40}&\factor{$<$ 0.01}&\factor{0.07}&\factor{0.03}&\factor{-2.40}&\factor{0.02}\\

Don't know&-0.05&0.05&-0.97&0.33&0.03&0.04&-0.83&0.41\\

\midrule 
\multicolumn{9}{@{}l}{\textbf{[Q10.2]: Knows app for symptom check  (baseline: yes)}}\\

\factor{No}&\factor{-0.21}&\factor{0.04}&\factor{-4.93}&\factor{$<$ 0.01}&\factor{0.20}&\factor{0.03}&\factor{-6.69}&\factor{$<$ 0.01}\\

Don't know&-0.01&0.06&-0.20&0.84&\factor{0.08}&\factor{0.04}&\factor{-2.10}&\factor{0.04}\\

\midrule \multicolumn{9}{@{}l}{\textbf{[Q10.3]: Knows app for quarantine enforcement (baseline: yes)}}\\

\factor{No}&\factor{-0.50}&\factor{0.05}&\factor{10.76}&\factor{$<$ 0.01}&\factor{0.23}&\factor{0.03}&\factor{-7.19}&\factor{$<$ 0.01}\\

\factor{Don't know}&\factor{-0.56}&\factor{0.06}&\factor{-9.02}&\factor{$<$ 0.01}&\factor{0.10}&\factor{0.04}&\factor{-2.40}&\factor{0.02}\\

\midrule \multicolumn{9}{@{}l}{\textbf{[Q10.4]: Knows app for contact tracing (baseline: yes)}}\\

No&\factor{-0.17}&\factor{0.04}&\factor{-4.12}&\factor{$<$ 0.01}&0.02&0.03&-0.67&0.51\\

Don't know&\factor{-0.19}&\factor{0.05}&\factor{-3.51}&\factor{$<$ 0.01}&0.03&0.04&0.87&0.39\\

\midrule \multicolumn{9}{@{}l}{\textbf{[Q10.5]: Knows app for health certificate (baseline: yes)}}\\

\factor{No}&\factor{-0.58}&\factor{0.04}&\factor{13.30}&\factor{$<$ 0.01}&\factor{0.24}&\factor{0.03}&\factor{-7.50}&\factor{$<$ 0.01}\\

\factor{Don't know}&\factor{-0.19}&\factor{0.06}&\factor{-3.39}&\factor{$<$ 0.01}&\factor{0.10}&\factor{0.04}&\factor{-2.34}&\factor{0.02}\\

\midrule \multicolumn{9}{@{}l}{\textbf{[Q18]: Uses any COVID-19 app (baseline: yes)}}\\

No&&&&&\factor{0.81}&\factor{0.09}&\factor{-8.76}&\factor{$<$ 0.01}\\

Don't know&&&&&\factor{0.67}&\factor{0.16}&\factor{-4.13}&\factor{$<$ 0.01}\\

\midrule \multicolumn{9}{@{}l}{\textbf{[Q28]: Rate measures (baseline: too strict)}}\\

\factor{About right}&\factor{0.31}&\factor{0.11}&\factor{2.84}&\factor{$<$ 0.01}&\factor{0.40}&\factor{0.05}&\factor{7.97}&\factor{$<$ 0.01}\\

\factorrounds{About right:R2}&-&-&-&-&\factorrounds{0.22}&\factorrounds{0.07}&\factorrounds{3.33}&\factorrounds{$<$ 0.01}\\

\factor{Too lenient}&\factor{0.45}&\factor{0.12}&\factor{3.82}&\factor{$<$ 0.01}&\factor{0.73}&\factor{0.06}&\factor{12.03}&\factor{$<$ 0.01}\\

\factorrounds{Too lenient:R2}&-&-&-&-&\factorrounds{-0.17}&\factorrounds{0.08}&\factorrounds{-2.18}&\factorrounds{0.03}\\

\factorrounds{Too lenient:R3}&-&-&-&-&\factorrounds{-0.24}&\factorrounds{0.08}&\factorrounds{-3.16}&\factorrounds{$<$ 0.01}\\

\midrule 
\multicolumn{9}{@{}l}{\textbf{[Q29.1]: Opinion of health authorities (baseline: neutral)}}\\

\factor{Favorable}&\factor{0.39}&\factor{0.09}&\factor{4.30}&\factor{$<$ 0.01}&\factor{0.16}&\factor{0.03}&\factor{5.77}&\factor{$<$ 0.01}\\

Unfavorable&\factor{0.41}&\factor{0.13}&\factor{3.14}&\factor{$<$ 0.01}&0.03&0.04&-0.69&0.49\\

\factorrounds{Unfavorable:R3}&\factorrounds{-0.77}&\factorrounds{0.15}&\factorrounds{-5.02}&\factorrounds{$<$ 0.01}&-&-&-&-\\

\midrule
\multicolumn{9}{@{}l}{\textbf{[Q29.2]: Opinion of law enforcement (baseline: neutral)}}\\

\factor{Favorable}&\factor{0.16}&\factor{0.03}&\factor{4.91}&\factor{$<$ 0.01}&\factor{0.16}&\factor{0.02}&\factor{6.72}&\factor{$<$ 0.01}\\

\factor{Unfavorable}&\factor{0.22}&\factor{0.04}&\factor{5.01}&\factor{$<$ 0.01}&\factor{0.22}&\factor{0.04}&\factor{-6.11}&\factor{$<$ 0.01}\\

\pagebreak 
\multicolumn{9}{@{}l}{\textbf{[Q29.3]: Opinion of research institutions (baseline: neutral)}}\\

Favorable&&&&&0.01&0.03&0.20&0.85\\

Unfavorable&&&&&\factor{0.11}&\factor{0.04}&\factor{-2.50}&\factor{0.01}\\

\midrule 
\multicolumn{9}{@{}l}{\textbf{[Q29.4]: Opinion of private companies (baseline: neutral)}}\\

\factor{Favorable}&\factor{0.30}&\factor{0.03}&\factor{9.04}&\factor{$<$ 0.01}&\factor{0.19}&\factor{0.02}&\factor{7.74}&\factor{$<$ 0.01}\\

Unfavorable&\factor{-0.11}&\factor{0.04}&\factor{-2.64}&\factor{$<$ 0.01}&0.05&0.03&-1.57&0.12\\

\midrule \multicolumn{9}{@{}l}{\textbf{[Q29.5]: Opinion of federal government (baseline: neutral)}}\\

Favorable&0.06&0.10&0.61&0.54&0.06&0.04&-1.57&0.12\\

Unfavorable&0.09&0.09&1.00&0.32&\factor{0.19}&\factor{0.04}&\factor{-4.77}&\factor{$<$ 0.01}\\

\factorrounds{Unfavorable:R3}&\factorrounds{-0.40}&\factorrounds{0.11}&\factorrounds{-3.48}&\factorrounds{$<$ 0.01}&-&-&-&-\\

\midrule \multicolumn{9}{@{}l}{\textbf{[Q29.6]: Opinion of state government (baseline: neutral)}}\\

\factor{Favorable}&\factor{0.33}&\factor{0.09}&\factor{3.46}&\factor{$<$ 0.01}&\factor{0.14}&\factor{0.03}&\factor{3.97}&\factor{$<$ 0.01}\\

\factorrounds{Favorable:R3}&\factorrounds{-0.39}&\factorrounds{0.11}&\factorrounds{-3.48}&\factorrounds{$<$ 0.01}&-&-&-&-\\

Unfavorable&-0.14&0.11&-1.26&0.21&0.01&0.04&-0.24&0.81\\

\midrule \multicolumn{9}{@{}l}{\textbf{Gender (baseline: Female)}}\\

Male&0.11&0.07&1.64&0.10&\factor{0.10}&\factor{0.02}&\factor{4.73}&\factor{$<$ 0.01}\\

\midrule \multicolumn{9}{@{}l}{\textbf{Region (US)}}\\

North east&0.20&0.11&1.88&0.06&&&&\\

South&\factor{0.27}&\factor{0.09}&\factor{2.98}&\factor{$<$ 0.01}&&&&\\

West&\factor{0.22}&\factor{0.10}&\factor{2.19}&\factor{0.03}&&&&\\

\midrule \multicolumn{9}{@{}l}{\textbf{Education (baseline: Less than high school)}}\\

\factor{High school or associate degree}&\factor{-0.62}&\factor{0.14}&\factor{-4.53}&\factor{$<$ 0.01}&\factor{0.36}&\factor{0.08}&\factor{4.24}&\factor{$<$ 0.01}\\

\factorrounds{High school or associate degree:R2}&\factorrounds{0.37}&\factorrounds{0.17}&\factorrounds{2.17}&\factorrounds{-0.03}&\factorrounds{-0.44}&\factorrounds{0.11}&\factorrounds{-4.04}&\factorrounds{$<$ 0.01}\\

\factorrounds{High school or associate degree:R3}&\factorrounds{0.52}&\factorrounds{0.17}&\factorrounds{3.14}&\factorrounds{$<$ 0.01}&-&\factorrounds{-}&\factorrounds{-}&\factorrounds{-}\\

Bachelor's degree&\factor{-0.62}&\factor{0.13}&\factor{-4.66}&\factor{$<$ 0.01}&0.15&0.09&1.66&0.10\\

\factorrounds{Bachelor's degree:R2}&\factorrounds{0.56}&\factorrounds{0.17}&\factorrounds{3.34}&\factorrounds{$<$ 0.01}&-&-&-&-\\

\factorrounds{Bachelor's degree:R3}&\factorrounds{0.60}&\factorrounds{0.16}&\factorrounds{3.66}&\factorrounds{$<$ 0.01}&-&-&-&-\\

Postgraduate Education&\factor{-0.81}&\factor{0.17}&\factor{-4.89}&\factor{$<$ 0.01}&&&&\\

\factorrounds{Postgraduate Education:R2}&\factorrounds{0.68}&\factorrounds{0.20}&\factorrounds{3.36}&\factorrounds{$<$ 0.01}&-&-&-&-\\

\factorrounds{Postgraduate Education:R3}&\factorrounds{1.18}&\factorrounds{0.20}&\factorrounds{6.04}&\factorrounds{$<$ 0.01}&-&-&-&-\\
\midrule \multicolumn{9}{@{}l}{\textbf{Age}}\\

Age&\factor{-0.01}&\factor{$<$ 0.01}&\factor{-4.39}&\factor{$<$ 0.01}&0.00&$<$ 0.01&-1.24&0.21\\

\factorrounds{Age:R3}&\factorrounds{-0.01}&\factorrounds{$<$ 0.02}&\factorrounds{-4.30}&\factorrounds{$<$ 0.01}&-&-&-&-\\

\midrule \multicolumn{9}{@{}l}{\textbf{[Q30] IUIPC}}\\

\factor{Collection}&\factor{-0.13}&\factor{0.03}&\factor{-3.98}&\factor{$<$ 0.01}&\factor{0.11}&\factor{0.03}&\factor{-4.21}&\factor{$<$ 0.01}\\

\factorrounds{Collection:R2}&\factorrounds{0.09}&\factorrounds{0.04}&\factorrounds{2.47}&\factorrounds{0.01}&-&-&-&-\\

\factorrounds{Collection:R3}&\factorrounds{0.10}&\factorrounds{0.04}&\factorrounds{2.70}&\factorrounds{$<$ 0.01}&-&-&-&-\\

Awareness&0.02&0.05&0.49&0.63&\factor{0.05}&\factor{0.02}&\factor{-2.37}&\factor{0.02}\\

\factorrounds{Awareness:R2}&\factorrounds{-0.16}&\factorrounds{0.06}&\factorrounds{-2.87}&\factorrounds{$<$ 0.01}&-&-&-&-\\

\factorrounds{Awareness:R3}&\factorrounds{-0.12}&\factorrounds{0.05}&\factorrounds{-2.18}&\factorrounds{0.03}&-&-&-&-\\

Control&-0.03&0.04&-0.60&0.55&0.01&0.02&0.35&0.72\\
\end{longtable}
\end{center}

%% file: tables/questionnaire_all_waves.tex
\section{Questionnaire}
\label{appendix:questionnaire}

\small

\textit{Please note: Questions marked with * indicate those for which we collected responses but did not report results in this paper for reasons of scope. Still, we report these questions here for the sake of completeness.}

\subsection*{Welcome Text}
The current situation with the novel coronavirus (SARS-CoV-2) and the disease it causes (COVID-19) has sparked an intense debate about the use of smartphone apps to better understand and contain the spread of the virus.
This study investigates how people perceive apps that promise to help fight the COVID-19 pandemic and what they expect from them.

\subsection*{Phone Use}

First we would like to ask you a few questions about the smartphone you mainly use.

\begin{enumerate}
\item[Q1:] Do you own a smartphone? 

[single choice: Yes / No / Prefer not to answer]

\item[Q2:] \textbf{If ``Yes'' in Q1:} What is your phone’s operating system? 

[single choice: Android/Google, iOS/Apple, Other (please specify:), Don’t know, Prefer not to answer]

\item[Q3:] \textbf{If ``Yes'' in Q1:} How satisfied are you with the following aspects of your smartphone? 

[array of single-choice questions for:
Battery life, Location accuracy (GPS), Camera quality, Speed (of apps)]

[answer options for each: Very satisfied, Satisfied, Neither satisfied nor dissatisfied, Dissatisfied, Prefer not to answer]

\end{enumerate}

\subsection*{Coronavirus Experience}

Now we would like to ask you some questions about your experience with the novel coronavirus.\\
This study uses the following terminology:
\begin{itemize}[noitemsep, nosep]
    \item ``coronavirus'': the novel coronavirus (SARS-CoV-2) that has caused a global pandemic in early 2020;
    \item ``COVID-19'': coronavirus disease 19, the respiratory disease caused by this virus;
    \item ``corona apps'': smartphone apps specifically developed to help limit the spread of the COVID-19 pandemic.
\end{itemize}

\begin{enumerate}

\item[Q4:] Are you or have you been infected with the novel coronavirus? 
[single choice; answer options: I was tested for coronavirus and at least one of the results was positive. / I was tested for coronavirus and all results were negative. / I was not tested for coronavirus and I do not think I have been infected. / I was not tested for coronavirus, but I suspect that I might have been infected. / Prefer not to answer]

\item[Q5:] Is there a person in your social circle who is or has been infected with the novel coronavirus? 

[single choice: Yes / No / Prefer not to answer]

\item[Q6:] Have you been quarantined or did you quarantine yourself because of coronavirus? 

[single choice: Yes / No / Prefer not to answer]

\item[Q7:] How concerned are you that you or someone you are close to will become infected with the coronavirus? 

[single choice: Not at all concerned / Slightly concerned / Somewhat concerned / Moderately concerned / Extremely concerned / Prefer not to answer]

\item[Q8:] To the best of your knowledge, is there a person at higher risk in your household, \ie, an older adult or a person of any age who has a serious underlying medical condition? 

[single choice: Yes / No / Prefer not to answer]

\item[Q9:] \textbf{Round 3 only:} Are you vaccinated against COVID-19?

[single choice; answer options: Yes, I am fully vaccinated / Yes, I am partially vaccinated / No, but I plan to get vaccinated in the future / No, and I do not plan to get vaccinated in the future / Prefer not to answer]

\end{enumerate}

\subsection*{Perception of Corona Apps}

\emph{Note: In Round 1, this section was displayed before the app scenarios. 
We moved these questions for Rounds 2 and 3 to reduce bias due to participants' prior exposure to the app scenarios.}

\begin{enumerate}
\item[Q10:] Do you know of any app recommended by the public authorities in (the United States | Germany) that can be used to ... 

[array of single-choice questions; answer options for each: Yes, No, Unsure, Prefer not to answer]
\begin{itemize}[noitemsep, nosep]
    \item ... get information about the novel coronavirus and its spread?
    \item ... check if you have coronavirus-related symptoms?
    \item ... enforce quarantine?
    \item ... identify people you have been in close contact with and alert them in case you tested positive for coronavirus?
    \item ... provide information about your health and needs to be shown if you want to visit a certain place?
\end{itemize}

\vspace{0.25cm}
\item[Q11:] In general, what do you consider positive aspects of smartphone apps to help limit the spread of the COVID-19 pandemic? [free text]

\vspace{0.25cm}
\item[Q12:] In general, what do you consider negative aspects of smartphone apps to help limit the spread of the COVID-19 pandemic? [free text]

\vspace{0.25cm}
\item[Q13:] \textbf{Round 3 only:} Imagine you could wish for a smartphone app to help contain the COVID-19 pandemic. What functionality would this app have? [free text]

\end{enumerate}

\subsection*{Introduction to App Scenarios}

In the following, you will be shown different apps to find out what kind of corona apps you would prefer to use. The presented apps are fictional and have different purposes and implement different functionalities. For each app, you will be asked a few questions. Please consider the app’s purpose and functionalities while answering the questions. Always assume that you are free to choose whether or not you install and use the app.

\subsection*{App Scenarios}

\textit{Note: This question group was displayed 10 times, only with different scenarios.}

\subsubsection*{Sample Scenario}

Imagine an app that provides information about the novel coronavirus and the disease it causes, COVID-19. The app uses your current or past location(s) and your COVID-19 infection status. In addition, the app collects general statistical data about you. This data is sent to research institutions when you request information and it will be stored for an unspecified period of time. It is possible that the app cannot download and display current information due to technical problems. In the future there may be other possible benefits of using this app. 

\begin{enumerate}

\item[Q14:] How likely are you to use this app on your smartphone? 

[single choice: 7-point scale with end points ``Very unlikely'' and ``Very likely'', plus ``Prefer not to answer'']

\item[Q15:] How many people in (Germany | the United States) do you expect to use this app on their smartphones? 

[single choice: 7-point scale with end points ``No one'' and ``Everyone'', plus ``Prefer not to answer'']

\item[Q16:] Please complete the following statement: Most people who are important to me think that I ... 

[single choice: 7-point scale with end points ``should not use this app'' and ``should use this app'', plus ``Prefer not to answer'']

\item[Q17:] How useful do you rate this app in helping contain the spread of the COVID-19 pandemic? 

[single choice: 7-point scale with end points ``Not at all useful'' and ``Very useful'', plus ``Prefer not to answer'']

\end{enumerate}
 
\subsection*{Use of Corona Apps}

\begin{enumerate}
\item[Q18:] \textbf{If ``Yes'' in Q1, for US in all rounds, for Germany only in Round 1:} Do you use any (kind of) corona app(s) on your smartphone? 

[single choice; answer options: Yes / No / \textbf{Rounds 2 and 3:} I had previously installed an app but I have since deleted it / Don't know / Prefer not to answer]

\item[Q19:] \textbf{If ``Yes'' in Q1, for Germany in rounds 2 and 3:} Do you use one of the following corona apps on your smartphone? 

[array of single-choice questions for: Corona-Warn-App / Corona-Datenspende (``Corona Data Donation'') / \textbf{Round 3 only:} Luca / Other app(s)]

[answer options for each: Yes / No / I had previously installed the app but I have since deleted it / Don't know / Prefer not to answer]

\item[Q20:] \textbf{If ``Other'' in Q19, for Germany in rounds 2 and 3:} You just stated to use or have used another corona app. Which app is / was it? [free text]

\item[Q21:] \textbf{If ``No'' in Q18 / Q19:} Why do you not use a corona app? [free text]

\item[Q22:] \textbf{If ``Yes'' or ``Previously installed'' in Q18 / 19:} Which corona app(s) do (\textbf{rounds 2 and 3:} or did) you use? [free text]

\item[Q23:] \textbf{If ``Previously installed'' in Q19; Rounds 2 and 3:} You just stated to have used a corona app but later deleted it. Why did you delete the corona app(s)? 
[free text]

\item[Q24:] \textbf{If ``Yes'' in Q21:}
How satisfied are you with the corona app(s) you currently use? [single choice: Very satisfied / Satisfied / Neither satisfied nor dissatisfied / Dissatisfied / Very dissatisfied / Prefer not to answer]

\end{enumerate}

\subsection*{Corona-Warn-App}

\emph{This section was only displayed to German participants in rounds 2 and 3 who had indicated to use the Corona-Warn-App in Q19.}

\begin{enumerate}
    \item[Q25*:] How satisfied are you with the German Corona-Warn-App? 
    
    [single choice: Very satisfied / Satisfied / Neither satisfied nor dissatisfied / Dissatisfied / Very dissatisfied / Prefer not to answer]
    
    \item[Q26*:] Did the Corona-Warn-App ever present you with ``Increased Risk'' (red warning as shown in the figure).
    
    [screenshot of said warning]
    
    [single choice: Yes / No / Don't know / Prefer not to answer]
    
    \item[Q27*:] Which of the following statements about the German Corona-Warn-App do you think are correct? The Corona-Warn-App ...
    [array of single-choice questions; answer options for each: True / False / Don't know / Prefer not to answer; correct answers shown here in \emph{italics}]
    
    \begin{itemize}[noitemsep, nosep]
        \item ... helps trace chains of infection by showing encounters with people who tested positive. \emph{(True)}
        \item ... uses location data from your smartphone. \emph{(False)}
        \item only collects data that do now allow for unique identification of the individual. \emph{(True)}
        \item automatically transmits the collected encounter data. \emph{(False)}
        \item ... transmits the collected encounter data to the Robert Koch Institute. \emph{(False)}
        \item ... stores the collected encounter data for an indefinite period of time. \emph{(False)}
        \item ... needs to be installed in order to freely travel between German states. \emph{(False)}
        \item ... could possibly not warn you even though you had close contact with an infected person who also uses the app. \emph{(True)}
    \end{itemize}
    
\end{enumerate}
 
\subsection*{Trust in Institutions}

\begin{enumerate}
\item[Q28:] How do you rate the measures taken in your area to fight the COVID-19 pandemic? 

[single choice: Way too strict / Too strict / About right / Too lenient / Way too lenient / Prefer not to answer]

\item[Q29:] What is your overall opinion of the following institutions in the COVID-19 pandemic? 

[array of single-choice questions for: Health authorities / Law enforcement / Research institutions / Private companies / Federal government / State government answer 

[answer options for each: Very favorable, Mostly favorable, Neither favorable nor unfavorable, Mostly unfavorable, Very unfavorable, Prefer not to answer]



\end{enumerate}

\subsection*{Individual Privacy Concerns}

\begin{enumerate}
\item[Q30:] Please indicate to what extent you agree with each of following statements.

[IUIPC constructs for Control, Awareness (of Privacy Practices), and Collection~\cite{malhotra_iuipc_2004}]
\end{enumerate}

%% file: tables/codebooks.tex
\section{Codebooks}
\label{app:codebooks}


\begin{table*}[ht]
\centering
\small
\caption[Combined codebooks for reasons not to use a COVID-19 app (Q21, ``WNA'' for ``why no app'') and negative aspects of such apps (Q12, ``NEG'').]{\label{tab:codebook-wna-negative}
Combined codebooks for reasons not to use a COVID-19 app (Q21, ``WNA'' for ``why no app'') and negative aspects of such apps (Q12, ``NEG''). Codes with ``--'' in the WNA / NEG column were not part of the respective codebook.
}

\setlength{\tabcolsep}{2pt}
\begin{tabular*}{\textwidth}{@{}>{\bfseries}clccl}

\toprule
&  \textbf{Code} & \textbf{WNA} & \textbf{NEG} & \textbf{Examples} \\
\midrule 
\addlinespace[.5\defaultaddspace]
\multirow{4}{*}{\rotatebox[origin=c]{90}{\parbox[c]{40pt}{\centering Privacy~/ Security}}}
& (govt.) surveillance & & & ``Big Brother'', ``used for more than just corona'' \\
\addlinespace[.75\defaultaddspace]
& privacy       & & & ``leak information'', ``invasive''                        \\
\addlinespace[.75\defaultaddspace]
& security      & & & ``too insecure'', ``could be hacked''                    \\
\addlinespace[.75\defaultaddspace]
\midrule\multirow{4}{*}{\rotatebox[origin=c]{90}{\parbox{30pt}{\centering Avail\-abil\-i\-ty}}}
& no app            & & -- & ``didn't know there's such an app'', ``no app yet''                \\
& no suitable app          & & -- & ``don't know a good one'', ``no approved [...] app''                \\
& lack of information      & & -- & ``don't know how to use it'', ``not familiar''                  \\
& no phone         & & -- & ``I do not have a smartphone'' \\

\midrule\multirow{7}{*}{\rotatebox[origin=c]{90}{\parbox{60pt}{\centering Psychological~/ Societal}}}
& user base         & & & ``not enough users'', ``need to be used correctly''       \\
& discrimination        & & & ``division into good and evil", ``stigmatization''    \\
& disinformation         & & & ``more fake news'', ``government may hide the truth''    \\
& anxiety           & & & ``would make me more nervous'', ``freaks people out''         \\
& false sense of protection     & & & ``blind faith in the [...] app''      \\
& autonomy         & & & ``loss of freedom'', ``civil rights'', ``unconstitutional''    \\
& not mandatory        & & & ``because I don't need to install it'', ``not compulsory''     \\

\midrule\multirow{5}{*}{\rotatebox[origin=c]{90}{Technical}}
& phone use         & & -- & ``do not carry phone all the time'', ``no mobile data''           \\
& not supported         & & -- & ``my phone is too old'', ``does not support new apps''\\
& technical side effects            & & & ``drains battery too much'', ``use[s] up the memory'' \\
& malfunctions            & & & ``not reliable'', ``not [...] accurate'', ``false positives''   \\
& inconvenient          & & & ``too complicated'', ``need to sign in every day''    \\

\midrule\multirow{7}{*}{\rotatebox[origin=c]{90}{Unnecessary}}
& state of the pandemic           & & & ``too late'', ``low risk region''           \\
& personal behavior             & & -- & ``I rarely go out'', ``my range of activities is small'' \\
& other information sources       & & & ``I watch TV'', ``I go to websites for information''    \\
& vaccinated        & & & ``I'm fully vaccinated'', ``I already had corona''     \\
& other measures          & & & ``social distancing'', ``I wear a mask''    \\
& coronaskeptic        & & & ``corona is fake'', ``just a flu'', ``conspiracy''     \\
& unnecessary (general)        & & & ``useless'', ``a waste of time'', ``I don't need it''  \\

\midrule\multirow{4}{*}{\rotatebox[origin=c]{90}{Other}}
& don't want            & & -- & ``I just don't want to'', ``no'', ``stupid''  \\
& generic negative            & -- & & ``everything'', ``yes'', ``many''       \\
& other          & & & ``too much information'', ``I do not trust them''    \\
& none          & -- & & ``I don't see any'', ``nothing really''         \\

\midrule\multirow{4}{*}{\rotatebox[origin=c]{90}{\parbox{40pt}{\centering Non-answers}}}
& don't know          & & & ``don't know'', ``not sure'', ``hard to tell'' \\
& positive           & -- & & ``help[s] keep people safe'', ``practical'', ``very good''\\
& unclear           & & & ``data problem'', ``Facebook'', ``more abundant''          \\
& no answer          & & & ``./.'', ``-'', ``vwedv'', ``Don't answer at the present time''  \\

\bottomrule
\end{tabular*}
\end{table*}


\begin{table*}[htbp]

\centering
\small
\caption[Codebook for positive aspects of COVID-19 apps (Q11)]{\label{tab:codebook-positive}
Codebook for positive aspects of COVID-19 apps (Q11)
}

\begin{tabular*}{\textwidth}{@{}>{\bfseries}cll}
\toprule
&  \textbf{Code} & \textbf{Examples} \\

\midrule\multirow{6}{*}{\rotatebox[origin=c]{90}{App Purposes}}

& contact tracing      & ``to trace affected people''                              \\
& symptom check        & ``help identify symptoms''                                \\
& quarantine           & ``enforcing quarantine''                                  \\
& information          & ``know the situation around''                             \\
& health certificate   & ``health QR code'', ``you can show your status''          \\
& movement             & ``record movement tracks'', ``travel route''              \\

\midrule
\rotatebox[origin=c]{90}{\parbox{30pt}{\centering Avail\-a\-bil\-i\-ty}}
& availability         & ``everyone has a smartphone'', ``convenient''             \\
\midrule
\addlinespace[1.25\defaultaddspace]
\multirow{2}{*}{\rotatebox[origin=c]{90}{\parbox{30pt}{\centering Psycho\-logical}}}
& sense of protection  & ``makes one feel safer'', ``relieve stress''              \\
\addlinespace[1.25\defaultaddspace]
& awareness            & ``remind us'', ``take it more seriously''                 \\
\addlinespace[1.25\defaultaddspace]

\midrule
\addlinespace[.25\defaultaddspace]
\multirow{4.5}{*}{\rotatebox[origin=c]{90}{\parbox{40pt}{\centering Pandemic Control}}}
& measures             & ``assess the risk levels [...] and act accordingly''      \\
\addlinespace[.25\defaultaddspace]
& infection detection  & ``knowledge of hot spots'', ``identify infected people''  \\
\addlinespace[.25\defaultaddspace]
& infection prevention & ``reduce the risk of infection'', ``keep people safe''    \\
\addlinespace[.25\defaultaddspace]
& generic pandemic     &  ``limit the spread'', ``control the outbreak''           \\
\addlinespace[.25\defaultaddspace]

\midrule
\multirow{3}{*}{\rotatebox[origin=c]{90}{Other}}
& generic positive     & ``very good'', ``it helps'', ``many''                     \\
& other                & ``big data'', ``publicity'', ``no more home office''      \\
& none                 & ``none'', ``not really'', ``nothing''                     \\

\midrule
\multirow{4}{*}{\rotatebox[origin=c]{90}{\parbox{35pt}{\centering Non-answers}}}
& don't know           & ``don't know'', ``hard to say'', ``not sure''             \\
& negative             & ``virus gets on phone'', ``invasion of privacy''          \\
& unclear              & ``Ues it impacts'', ``Japanese style'', ``ccorona''       \\
& no answer            & ``jhgfkjfkuz'', ``Tv'', ``na'', ``??''                    \\

\bottomrule
\end{tabular*}
\end{table*}


\begin{table*}[htbp]

\centering
\small
\caption[Codebook for participants' ``ideal'' COVID-19 app (Q13)]{\label{tab:codebook-idealapp}
Codebook for participants' ``ideal'' COVID-19 app (Q13)
}

\begin{tabularx}{\textwidth}{@{}>{\bfseries}clX}
\toprule
&  \textbf{Code} & \textbf{Examples} \\

\midrule\multirow{6}{*}{\rotatebox[origin=c]{90}{Information}}

& vaccination / test sites   & ``book [appointments for] corona tests and vaccinations'', ``find spare doses of the vaccine''          \\
& regulations      & ``show exactly what is necessary where (test, appointment, etc.)'', ``corona restrictions''                            \\
& hotspots     &  ``notification of outbreak'', ``show what areas are infected''           \\
& information          & ``CDC updates'', ``information on the spread in my area''                             \\
\midrule

\multirow{11}{*}{\rotatebox[origin=c]{90}{Pandemic Control}}

& contact tracing      & ``to trace affected people'', ``determine contacts'', ``tell me when/where I met someone positive''                             \\
& contact diary      & ``the ability to enter recent contacts quickly and accurately'', ``contact diary''                              \\
& quarantine           & ``keep infected people away from others'', ``notification if someone needs to be quarantined''                                  \\
& movement             & ``show how many people are currently at which location'', ``track people's location''              \\
& infection prevention & ``reduce the risk of infection'', ``keep people safe''    \\
& real-time detection  & ``automatically show if friends are positive'', ``alert me if an infected person is moving towards me'' \\
& infection detection  & ``identify infected individuals'' \\

\midrule

\multirow{5}{*}{\rotatebox[origin=c]{90}{Certificates}}

& vaccine passport   & ``carry vaccine passport'', ``vaccination information''          \\
& test results   & ``proof of tests'', ``functionality that shows a negative test'', ``entering the test result''          \\
& health certificate   & ``QR code of my data'', ``a way to test you before you access a building [that] requires it''          \\

\midrule
\addlinespace[1.25\defaultaddspace]
\multirow[c]{2}{*}{\rotatebox[origin=c]{90}{\parbox[c]{36pt}{\centering Medical Support}}}
& treatment      & ``treat the virus'', ``shoot out the vaccine'' \\
\addlinespace[1.25\defaultaddspace]
& symptom check        & ``symptom tracking'', ``sense of fever'', ``testing'', ``corona test'' \\
\addlinespace[1.25\defaultaddspace]

\midrule
\multirow{5}{*}{\rotatebox[origin=c]{90}{Other}}
& generic positive     & ``yes'', ``great'', ``stop it'', ``quality'', ``fight Covid''                     \\
& property     & ``privacy'', ``fast'', ``easy to use'', ``everyone has to use it'', ``a block button''                     \\
& existing app     & ``Corona-Warn-App'', ``Instagram'', ``it already exists''                     \\
& other                & ``a drop dead answer to stop the pandemic'', ``makes people get the vaccine''      \\
& none                 & ``none'', ``I don't want any of these apps'', ``nothing'', ``no desire to be tracked''                     \\

\midrule
\multirow{4}{*}{\rotatebox[origin=c]{90}{\parbox{35pt}{\centering Non-answers}}}
& don't know           & ``don't know'', ``hard to say'', ``not sure''             \\
& unclear              & ``Digitarer'', ``covid'', ``neutral'', ``it may'', ``good riddance ya ya know'', ``innovate worry''       \\
& no answer            & ``gtzhui'', ``na'', ``??'', ``don't care'', ``not relevant'', ``not interested''                    \\

\bottomrule
\end{tabularx}
\end{table*}

%% file: tables/deus123_descriptives/Q14_positive_mean.tex
\begin{table}[h]
\caption{Mean response values ($\pm$ sd) for Q14 (willingness to use apps).}
\centering
\begin{tabular}{lcccccc}
\toprule
\textbf{App Purpose} &\multicolumn{3}{c}{\textbf{Germany}} & \multicolumn{3}{c}{\textbf{United States}} \\

\cmidrule{2-4}
\cmidrule{5-7}
& R1 & R2 & R3 & R1 & R2 & R3 \\
\midrule
Contact tracing & $3.58 \pm 2.09$ & $3.31 \pm 2.09$ & $3.47 \pm 2.09$ & $3.35 \pm 2.13$ & $3.34 \pm 2.19$ & $3.16 \pm 2.19$\\ 
Health certificate & $3.02 \pm 1.99$ & $2.75 \pm 1.96$ & $3.33 \pm 2.07$ & $3.06 \pm 2.09$ & $3.07 \pm 2.14$ & $3.02 \pm 2.15$\\ 
Information & $3.29 \pm 2.01$ & $2.93 \pm 1.99$ & $3.16 \pm 2.01$ & $3.03 \pm 2.05$ & $3.09 \pm 2.13$ & $2.98 \pm 2.12$\\ 
Quarantine Enf. & $3.05 \pm 2.03$ & $2.71 \pm 1.97$ & $3.03 \pm 2.03$ & $2.95 \pm 2.08$ & $2.95 \pm 2.13$ & $2.89 \pm 2.15$\\ 
Symptom check & $3.32 \pm 2.04$ & $3.01 \pm 1.99$ & $3.18 \pm 1.99$ & $3.20 \pm 2.11$ & $3.26 \pm 2.18$ & $3.09 \pm 2.16$\\ 
\bottomrule
\end{tabular}
\label{tab:Q9_mean}
\end{table}

%% file: tables/deus123_descriptives/Q17_utility_mean.tex
\begin{table}[h]
\caption{Mean response values ($\pm$ sd) for Q17 (perceived utility / usefulness).}
\centering
\begin{tabular}{lcccccc}
\toprule
\textbf{App Purpose} &\multicolumn{3}{c}{\textbf{Germany}} & \multicolumn{3}{c}{\textbf{United States}} \\

\cmidrule{2-4}
\cmidrule{5-7}
& R1 & R2 & R3 & R1 & R2 & R3 \\
\midrule
Contact tracing & $3.99 \pm 1.92$ & $3.70 \pm 1.93$ & $3.82 \pm 1.92$ & $3.93 \pm 1.98$ & $3.81 \pm 2.01$ & $3.53 \pm 2.04$\\ 
Health certificate & $3.51 \pm 1.86$ & $3.18 \pm 1.85$ & $3.61 \pm 1.91$ & $3.50 \pm 1.99$ & $3.47 \pm 2.00$ & $3.30 \pm 2.02$\\ 
Information & $3.62 \pm 1.84$ & $3.23 \pm 1.86$ & $3.41 \pm 1.86$ & $3.45 \pm 1.95$ & $3.47 \pm 1.98$ & $3.25 \pm 1.99$\\ 
Quarantine Enf. & $3.63 \pm 1.94$ & $3.27 \pm 1.92$ & $3.50 \pm 1.96$ & $3.57 \pm 1.99$ & $3.51 \pm 2.01$ & $3.27 \pm 2.04$\\ 
Symptom check & $3.69 \pm 1.87$ & $3.33 \pm 1.86$ & $3.51 \pm 1.87$ & $3.67 \pm 1.96$ & $3.67 \pm 1.99$ & $3.39 \pm 2.01$\\ 
\bottomrule
\end{tabular}
\label{tab:Q12_mean}
\end{table}

%% file: tables/deus123_descriptives/interactionplots.tex
\section{Interaction Plots}
\label{sec:interactionplots}

\begin{figure}[ht]
\centering
    \begin{subfigure}[t]{0.49\columnwidth}
        \includegraphics[width=\textwidth]{figures/interaction_purpose_immunity_q9.pdf}
    \end{subfigure}
    \begin{subfigure}[t]{0.49\columnwidth}
        \includegraphics[width=\textwidth]{figures/interaction_purpose_contacttracing_q9.pdf}
    \end{subfigure}
    \hfill
        \begin{subfigure}[t]{0.49\columnwidth}
        \includegraphics[width=\textwidth]{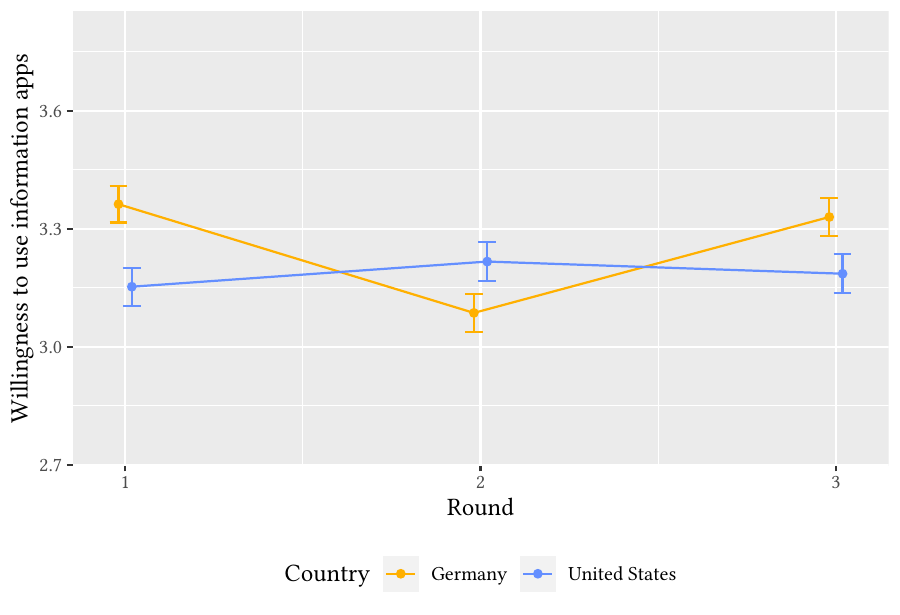}
    \end{subfigure}
        \begin{subfigure}[t]{0.49\columnwidth}
        \includegraphics[width=\textwidth]{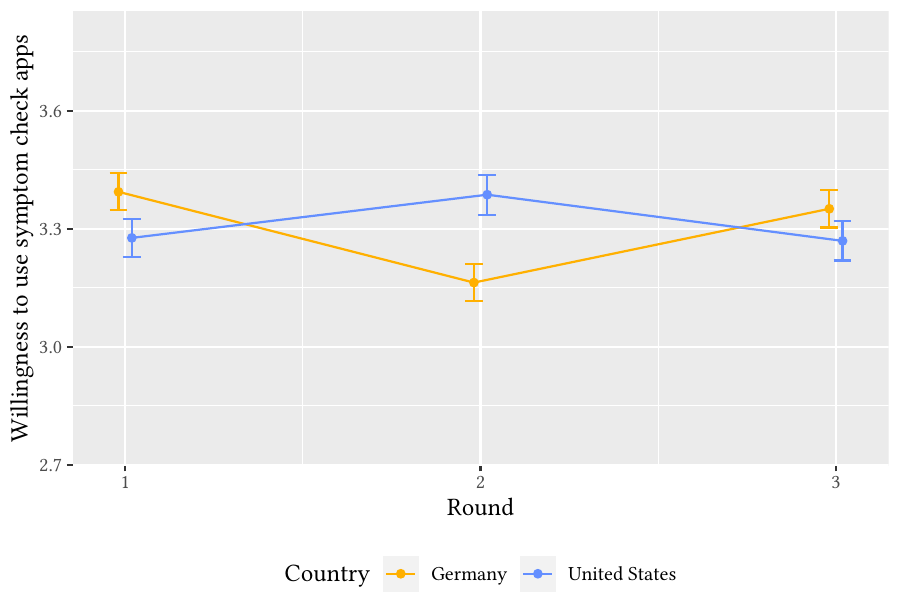}
    \end{subfigure}
    \hfill
        \begin{subfigure}[t]{0.49\columnwidth}
        \includegraphics[width=\textwidth]{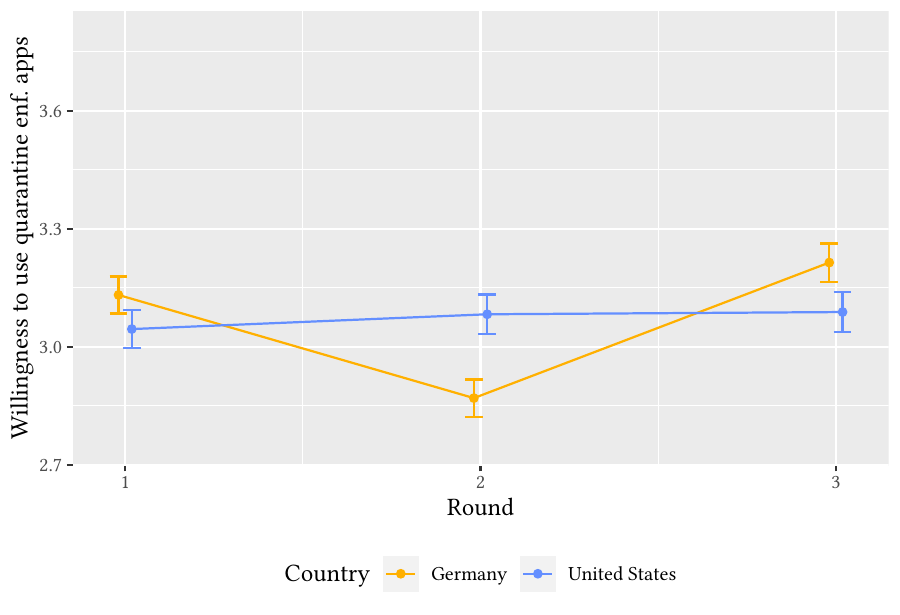}
    \end{subfigure}
    \vskip\baselineskip
    \caption{Willingness to use COVID-19 apps in interaction with the survey round (\ie, over time), differentiated by app purposes: health certificate (top left), contact tracing (top right), information (mid left), symptom check (mid right), and quarantine enforcement (bottom).}
    \label{fig:interaction_allpurposes_usde}
\end{figure}

%% file: tables/deus123_descriptives/interactionplots_concerns.tex
\begin{figure}[tb]
\centering
    \begin{subfigure}[b]{0.49\columnwidth}
        \includegraphics[width=\textwidth]{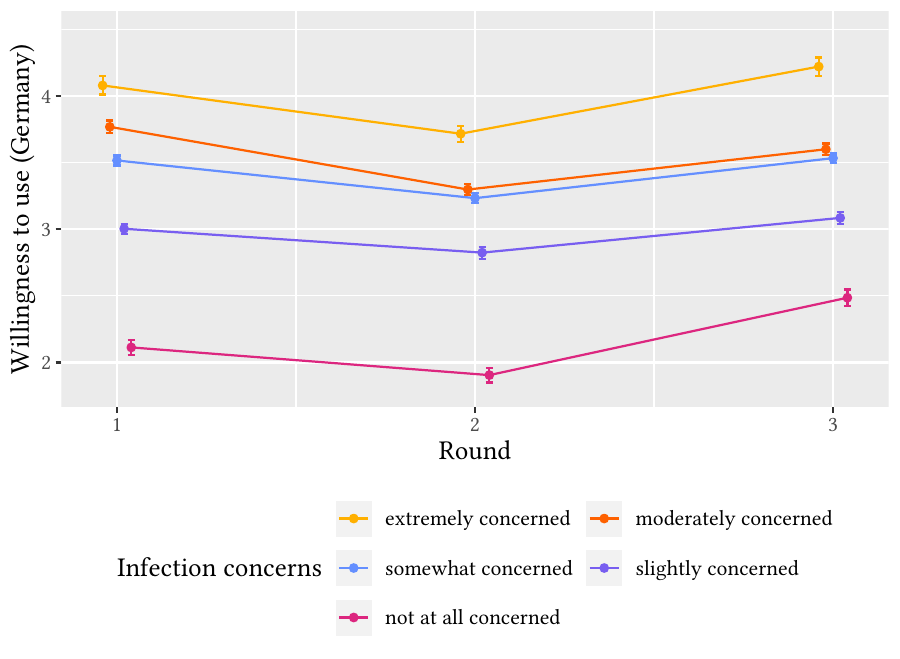}
    \end{subfigure}
        \begin{subfigure}[b]{0.49\columnwidth}
        \includegraphics[width=\textwidth]{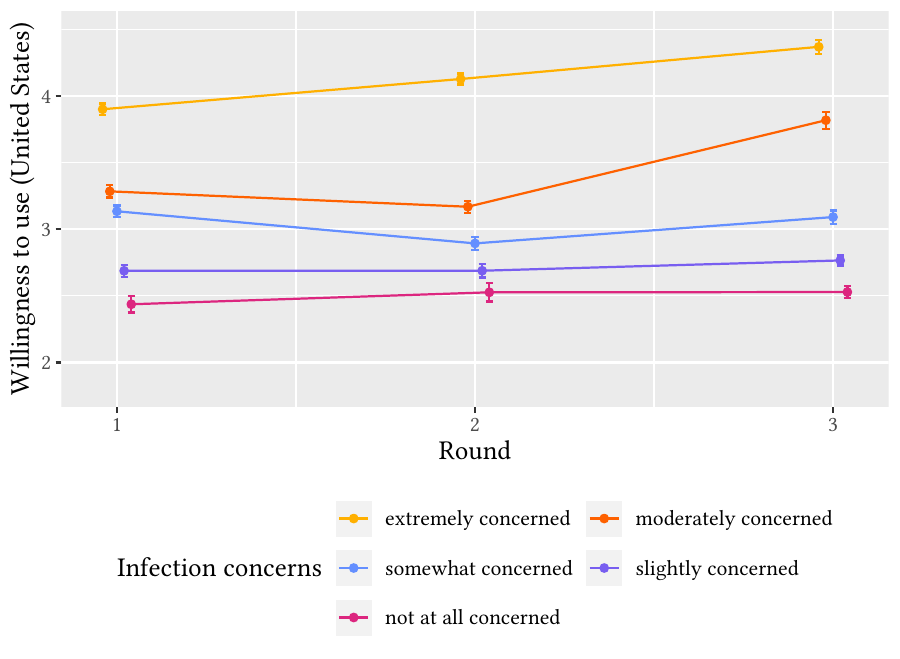}
    \end{subfigure}
    \caption{Willingness to use COVID-19 apps in Germany (left) and the United States (right) in relation to participants' concerns of becoming infected with the coronavirus (Q7) over time.}
    \label{fig:interaction_concerns_usde}
\end{figure}

%% file: tables/deus123_descriptives/interactionplots_stategovernment.tex
\begin{figure}[tb]
\centering
    \begin{subfigure}[b]{0.49\columnwidth}
        \includegraphics[width=\textwidth]{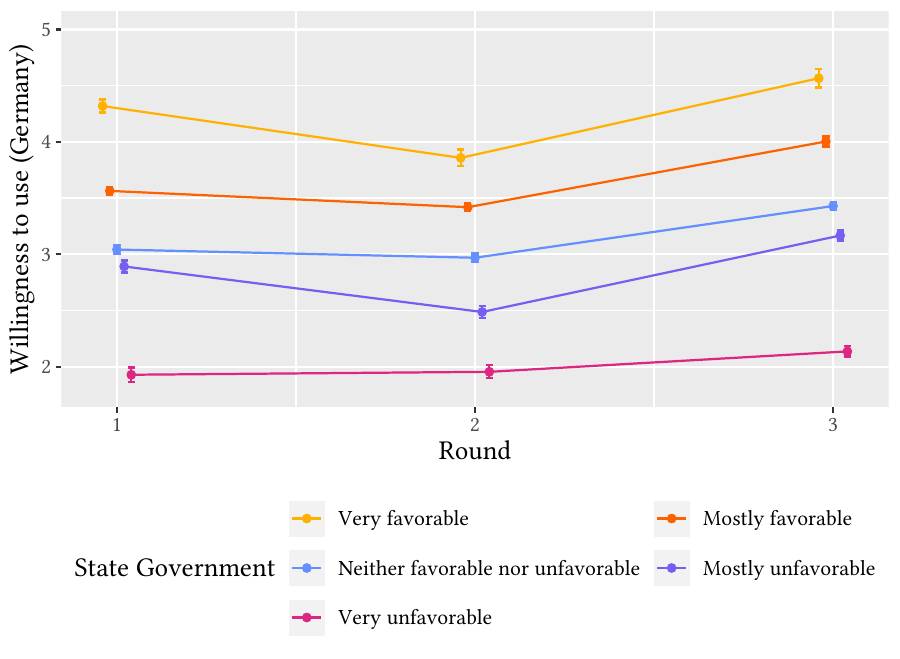}
    \end{subfigure}
        \begin{subfigure}[b]{0.49\columnwidth}
        \includegraphics[width=\textwidth]{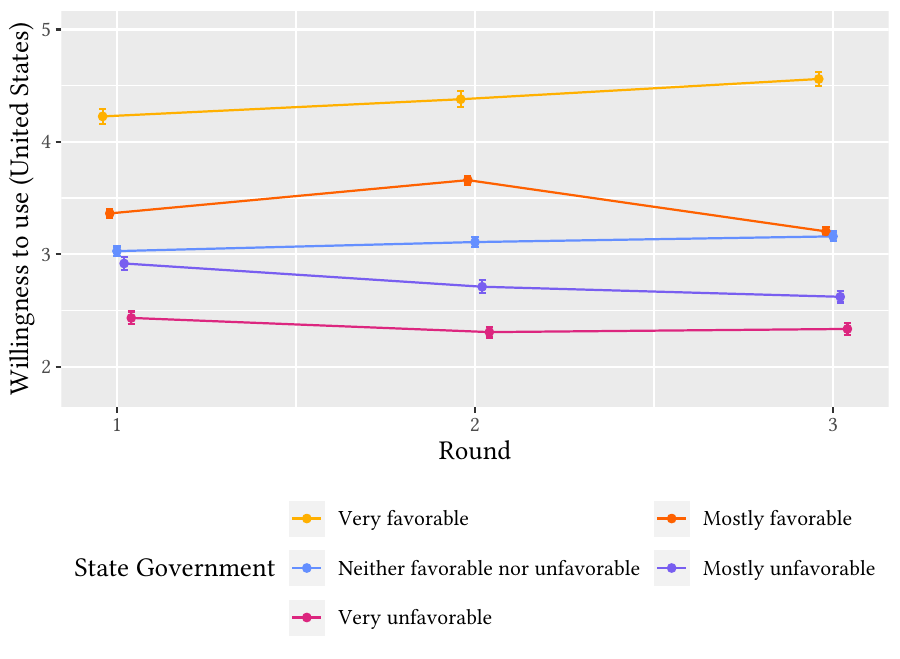}
    \end{subfigure}
    \caption{Willingness to use COVID-19 apps in Germany (left) and the United States (right) in relation to participants' overall opinion of the state government in the COVID-19 pandemic (Q29.6) over time.}
    \label{fig:interaction_stategovernment_usde}
\end{figure}